\documentclass{aa}  

\usepackage{graphicx}
\usepackage{txfonts}
\usepackage{lipsum}
\usepackage{subcaption}        
\usepackage{lscape}
\usepackage{placeins}
\usepackage{comment}
\usepackage{color}
\usepackage[dvipsnames]{xcolor}

\usepackage[colorlinks=true, pdfstartview=FitV, linkcolor=blue, citecolor=black, plainpages=false, pdfpagelabels=true, urlcolor=blue]{hyperref}

\def\ltsima{$\; \buildrel < \over \sim \;$}
\def\simlt{\lower.5ex\hbox{\ltsima}}
\def\gtsima{$\; \buildrel > \over \sim \;$}
\def\simgt{\lower.5ex\hbox{\gtsima}}
\def\arcsec{\hbox{$^{\prime\prime}$}}

\newcommand{\lsun}{$L_\odot$}
\newcommand{\msun}{$M_\odot$}

\begin{document}
\let\linenumbers\relax
\let\internallinenumbers\relax
\let\makeLineNumber\relax
\let\LineNumber\relax

   \title{Dissecting the dust distribution and polarization around two B213 young stellar objects with ALMA}

   \author{Asako Sato\inst{1},
        Ana\"elle Maury\inst{1,2,3},
        Josep M. Girart\inst{1,4},
        Andrea Bracco\inst{5,6},
        Patrick Hennebelle\inst{3},
        Qizhou Zhang\inst{7},
        \and Valeska Valdivia\inst{8,9}
        }

   \institute{Institut de Ciències de l’Espai (ICE-CSIC), Campus UAB, Can Magrans S/N, E-08193 Cerdanyola del Vallès, Catalonia, Spain\\
             \email{asako@ice.csic.es}
             \thanks{Shows the usage of elements in the author field}
            \and ICREA, Pg. Lluís Companys, 23, E-08010, Barcelona, Catalonia, Spain 
            \and Université Paris-Saclay, Université Paris Cité, CEA, CNRS, AIM, 91191, Gif-sur-Yvette, France
            \and Institut d’Estudis Espacials de Catalunya (IEEC), Campus del Baix Llobregat—UPC, Esteve Terradas 1, E-08860 Castelldefels, Catalonia, Spain
            \and INAF – Osservatorio Astrofisico di Arcetri, Largo E. Fermi 5, 50125 Firenze, Italy
            \and Laboratoire de Physique de l'Ecole Normale Sup\'erieure, ENS, Universit\'e PSL, CNRS, Sorbonne Universit\'e, Universit\'e de Paris, F-75005 Paris, France
            \and Center for Astrophysics, Harvard \& Smithsonian, 60 Garden Street, Cambridge, MA~02138, USA
            \and Liant, 5 Av. Jean Jaur\`es, 34600, B\'edarieux, France
            \and Department of Physics, Nagoya University, Furo-cho, Chikusa-ku, Nagoya, Aichi 464-8602, Japan
            }

  \abstract
   {The pristine stages of disk formation during the protostellar phase and the processes of dust evolution in young disks remain largely unconstrained. Dust thermal emission at millimeter wavelengths and its polarization offer key insights into the physical processes and spatial distribution of material at the envelope-disk interface, as well as on early dust evolution around solar-type protostars.}
   {We characterize the dust properties and polarization mechanisms from envelope to disk scales in two young stellar objects (YSOs), IRAS\,04166+2706 (K04166) and IRAS\,04169+2702 (K04169) embedded in the same Taurus filament. We explore the nature of the dust polarization across scales and investigate their stage of evolution.}
   {We present deep ($\sim$1.4\,mm and $\sim$3\,mm) polarimetric observations sampling the dust emission from 25\,au to 3000\,au using the Atacama Large
Millimeter/submillimeter Array (ALMA). 
   We modeled Stokes~$I$ emission to characterize disk and envelope contributions, while the polarization properties were analyzed to identify the dominant polarization mechanisms. Finally, in this work, we discuss the physical properties of both sources across scales.}
   {K04166 shows extended Stokes~$I$ and polarized emission tracing a tentative hourglass magnetic field morphology in its envelope. In the inner envelope and disk (<100\,au), the properties of the polarized emission change, suggesting the presence of a toroidal magnetic field around the disk or the presence of large grains in the inner envelope. K04169 exhibits compact Stokes~$I$ and polarized emission, consistent with self-scattering from the disk. Both disks are extremely compact, yet K04166 retains a substantial envelope while the one of K04169 is largely dissipated.
   }
   {Our multiscale ALMA polarimetric observations reveal a transition from magnetically aligned grains in envelopes to self-scattering in disks within the transition region of 20--50\,au.  These results provide important clues on dust grain growth and magnetic field morphology at the disk-envelope scales. 
   The two sources, separated by $<0.5$ pc and embedded in a common filament, display striking differences, indicating that K04166 is a young embedded object with a substantial envelope threaded by relatively organized magnetic fields. Meanwhile, K04169 is more evolved, likely to be a young T-Tauri star.
   However, in both disks, the presence of large grains already suggest a scenario of early dust evolution in disks of the Class\,0 stage.}


   \titlerunning{The polarization and dust emission in two B213 cores}
   \authorrunning{A. Sato et al.}
   \maketitle

\section{Introduction}

Magnetic fields are thought to play a crucial role in the star formation process. In particular, they could play a role in regulating the formation and fragmentation of dense cores, affecting angular momentum transport during collapse, and ultimately setting the conditions for disk formation through magnetic braking \citep{Shu1987, Hennebelle_2008A&A...477....9H, Machida_2008ApJ...676.1088M, Wurster2018, Maury_2022FrASS...9.9223M, Tsukamoto_2023ASPC..534..317T}. These questions are particularly relevant during the earliest protostellar stages. Class 0 protostars, with ages of only $\sim 10^{4}$ years, represent the phase when most of the final stellar mass is assembled, while young disks build up from the infalling material of the dense surrounding envelopes \citep{AWB2000, Looney2000, Jorgensen2007, Maury_2011A&A...535A..77M}. Thus, understanding the magnetic field topology down to the disk-forming scales in young protostellar cores is key to constraining models of core evolution and disk formation.

Observing magnetic fields in the early stages of the star formation process helps to improve our understanding of how they redistribute angular momentum, regulate disk formation, and drive outflows \citep{Maury_2022FrASS...9.9223M}.
Dust polarization has long been the primary observational tracer of magnetic fields in star-forming regions. It is widely accepted that nonspherical dust grains can align with local magnetic fields via radiative torque (RAT) alignment, leading to polarized thermal emission \citep{Lazarian_2007JQSRT.106..225L, Andersson_2015ARA&A..53..501A}. This technique has been successfully used from cloud scales down to dense cores, revealing mostly organized and dynamically relevant magnetic fields in the vicinity of solar-type low-mass protostars \citep{Girart_2006Sci...313..812G, Kwon_2019ApJ...879...25K, Takahashi_2019ApJ...872...70T, Huang_2024ApJ...963L..31H}.

However, recent ALMA observations have demonstrated that on disk scales, polarized emission at millimeter wavelengths can also arise from self-scattering of dust thermal emission \citep{Kataoka_2015ApJ...809...78K, Stephens_2017ApJ...851...55S}. This mechanism, due to the scattering of photons emitted by local thermal dust emission by the neighboring dust grains, occurs in high-density conditions in phases when grains approach sizes comparable to the observing wavelength \citep{Kataoka_2017ApJ...844L...5K, Yang_2017MNRAS.472..373Y}. 
While self-scattering complicates the interpretation of dust polarization as a magnetic-field tracer, especially in disks where it seems to dominate in many cases, it could serve as a useful tool for measuring the sizes of dust grains in the high-density regions of inner envelopes at the disk-envelope transition, in addition to placing constraints on the dust evolution processes from submicron solid particles toward millimeter-sized ones during the disk building stage. 
However, young disks are very compact \citep{Maury_2019A&A...621A..76M, Sheehan2022} and, as a result, there have not been many observations that have been capable of probing the scales from envelopes to disks in polarized emission. 
Additionally, the required sensitivity to analyze both the magnetic field topology in the envelopes and the polarization due to self-scattering in and around the disks is very challenging. A promising way forward is to jointly study the polarized emission at envelope and disk scales around young nearby protostars. By probing both the regions where magnetic fields are expected to dominate the dynamics and those where scattering may emerge, we can hope to directly trace how dust physics and magnetic field structure evolve during the earliest stages of star formation.

In Taurus, the B213 region ($d=156 \pm 1$ \,pc, \citealt{Gaia_Krolikowski_2021AJ....162..110K}) provides an ideal laboratory for such a study. Embedded within the L1495 complex, B213 is a $\sim 0.5$\,pc -long filament, forming stars since $\simgt $3\,Myr ago \citep{Soler2023}, has been extensively studied, and stands as a classic example of a star-forming filament \citep{Schmalzl2010, Palmeirim_2013A&A...550A..38P}. It consists of 35 velocity-coherent filaments traced by molecular line emission \citep[i.e., fibers;][]{Hacar_2013A&A...554A..55H} and also shows faint dust emission striations. These striations are perpendicular to the filament and generally oriented along the magnetic field direction of the cloud, as indicated by optical and far-infrared (FIR) polarization datasets \citep{Chapman_2011ApJ...741...21C, Soler2019, Chakali_2021ApJ...912L..27E}. 
B213 hosts several young stellar objects (YSOs) across a range of evolutionary stages, sharing a common origin and pristine environment but spanning different evolutionary stages, including two self-gravitating prestellar cores (Miz-2 and Miz-8b; \citealt{Mizuno_1994Natur.368..719M, Tatematsu_2004ApJ...606..333T}), one T-Tauri star (J04194148+2716070; \citealt{Davis_2010MNRAS.405..759D}), and two protostellar cores (IRAS\,04166+2706 and IRAS\,04169+2702; \citealt{Motte2001, Rebull_2010, Marsh_2016MNRAS.459..342M}) presented in Fig.~\ref{fig:NIKA-ALMA}.
The two protostars, IRAS\,04166+2706 (K04166, Class 0) and IRAS\,04169+2702 (K04169, Class I), have bolometric luminosities of 0.40\,\lsun and 1.46\,\lsun, and bolometric temperatures of 61\,K and 163\,K, respectively \citep{Motte2001, Ohashi_2023ApJ...951....8O}. 
An ALMA large program, eDisk, kinematically estimated their central stellar mass to be 0.27$\pm$0.12\,\msun for K04166 and 1.312$\pm$0.044\,\msun for K04169 \citep{Phuong_2025ApJ...992...18P, Han_2025ApJ...993..120H}.
These two sources are separated by $<0.5$ pc and show parallel molecular outflows along with the magnetic field showing an organized pattern perpendicular to the elongation of filament at $\sim$0.1\,pc scale \citep{Bontemps_1996A&A...311..858B, Tafalla_2004A&A...423L..21T,Narayanan_2012MNRAS.425.2641N, Goldsmith_2008ApJ...680..428G, Palmeirim_2013A&A...550A..38P}.
However, recent observations probing down to the core scales (< 0.1\,pc), have shown magnetic fields do not always remain aligned with the large-scale configuration: for example, the magnetic field line segments (hereafter, B-field vectors) toward K04166 and K04169 are found to be nearly orthogonal between the two cores \citep[at $\sim$ 14$\arcsec$ $\sim$ 0.01\,pc;][]{Chakali_2021ApJ...912L..27E}. 

These innate similarities coupled with striking differences make K04166 and K04169 excellent test cases for studying the magnetic field and dust evolution in systems born from similar environments. In this paper, we present new multiscale polarization observations of both objects, from $\sim$900\,au down to 25\,au. We analyze dust properties across wavelengths and physical scales to disentangle the contributions of magnetic alignment and scattering, investigate dust evolution within the common natal environment, and study the magnetic field morphology from envelope to disk scales.

The paper is organized as follows. In Sect.~\ref{sec_2}, we introduce ALMA observations and our data reduction process. Sect.~\ref{sec_3} presents Stokes~$I$ maps and identifications of disks and envelopes in the continuum images and visibilities, along with our investigation of the polarization mechanisms for both sources. In Sect.~\ref{sec_4}, we discuss the nature of the polarized emission in both sources and evolutionary differences among these "twins." Finally, we present our conclusions in Sect.~\ref{sec_5}.

\section{ALMA observations and data reductions}
\label{sec_2}
The two ALMA projects (\#2019.1.01209.S, \#2021.1.01420.S, PI A. Maury) analyzed here have provided deep full-polarization observations of B213 protostars at 95\,GHz and 218\,GHz, utilizing both the ACA 7-m array (the 7-m array) and the ALMA 12-m array (the 12-m array) across three array configurations. While full completion of the observational program was not achieved, the datasets obtained at each band probe contrasting spatial scales. The highest resolution dataset has enabled the most detailed mapping to date of dust polarized emission in these systems, sampling spatial scales two orders of magnitude finer than in prior studies \citep[see, e.g.,][]{Chakali_2021ApJ...912L..27E}.

This work focuses on the 3\,mm and 1.4\,mm dust continuum emission (the $^{12}$CO outflow maps are shown for discussion purposes in Fig.~\ref{fig:outflow_polari}, along with more details in Sect.~\ref{sub:discussion-K66}, yet molecular line emission is not analyzed in this study). The data reduction was conducted with CASA version 6.4.0 \citep{CASA_2022PASP..134k4501C}, employing three rounds of phase self-calibration \citep{Pearson1984} for each dataset to minimize image noise and enhance fidelity. This process typically improved the quality of the 1.4\,mm data from the 12-m array by a factor of approximately 7.0, for instance. Bright spectral lines were excluded during continuum extraction. The continuum maps were then produced using the \texttt{tclean} task in CASA.

We present the ALMA dataset for each of the two sources as a set of three maps sampling different scales and frequencies: 3mm (the 12-m array, Band\,3), 1mm-high (the 12-m array, Band\,6), and 1mm-combine (combined the 7-m and 12-m arrays, Band\,6). Imaging parameters are detailed in Tables\,\ref{obs_para} and \ref{tab:para-continuum}.

The debiased polarized intensity is defined as $P = (Q^2 + U^2 - \sigma^2_{QU})^{1/2}$, where $\sigma_{QU} = \sqrt{(\sigma_Q^2 + \sigma_U^2)/2}$, $\sigma_{Q}$ is Stokes $Q$ error, and $\sigma_{U}$ is the Stokes $U$ error \citep{Vaillancourt2006PASP..118.1340V, Hull2015JAI.....450005H}. 
We note that we call this the debiased polarized intensity the polarized intensity in this paper and polarization maps presented here are derived with the debiased polarized intensity.
We then calculate the fractional linear polarization of $P_{\mathrm{frac}} = P/I$. The position angle (P.A.) of polarization is also defined as $\chi = 1/2 \arctan(U/Q)$. To ensure robust positional sampling, we applied Super-Nyquist sampling, with pixels sampling the polarization direction (position angles) at approximately one-third of the major axis of the beam \citep{Hull_2020PASP..132i4501H}. We adopted a 3$\sigma$ cutoff in total intensity 2$\sigma$ cutoff in polarized intensity for polarization fraction mapping and a 3$\sigma$ threshold on the polarized intensity for the polarization vector maps.

\section{Results}
\label{sec_3}
\subsection{Stokes~$I$ dust continuum emission: an overview}

\begin{figure*}[h!]
    \centering
    \includegraphics [width=17.5cm]{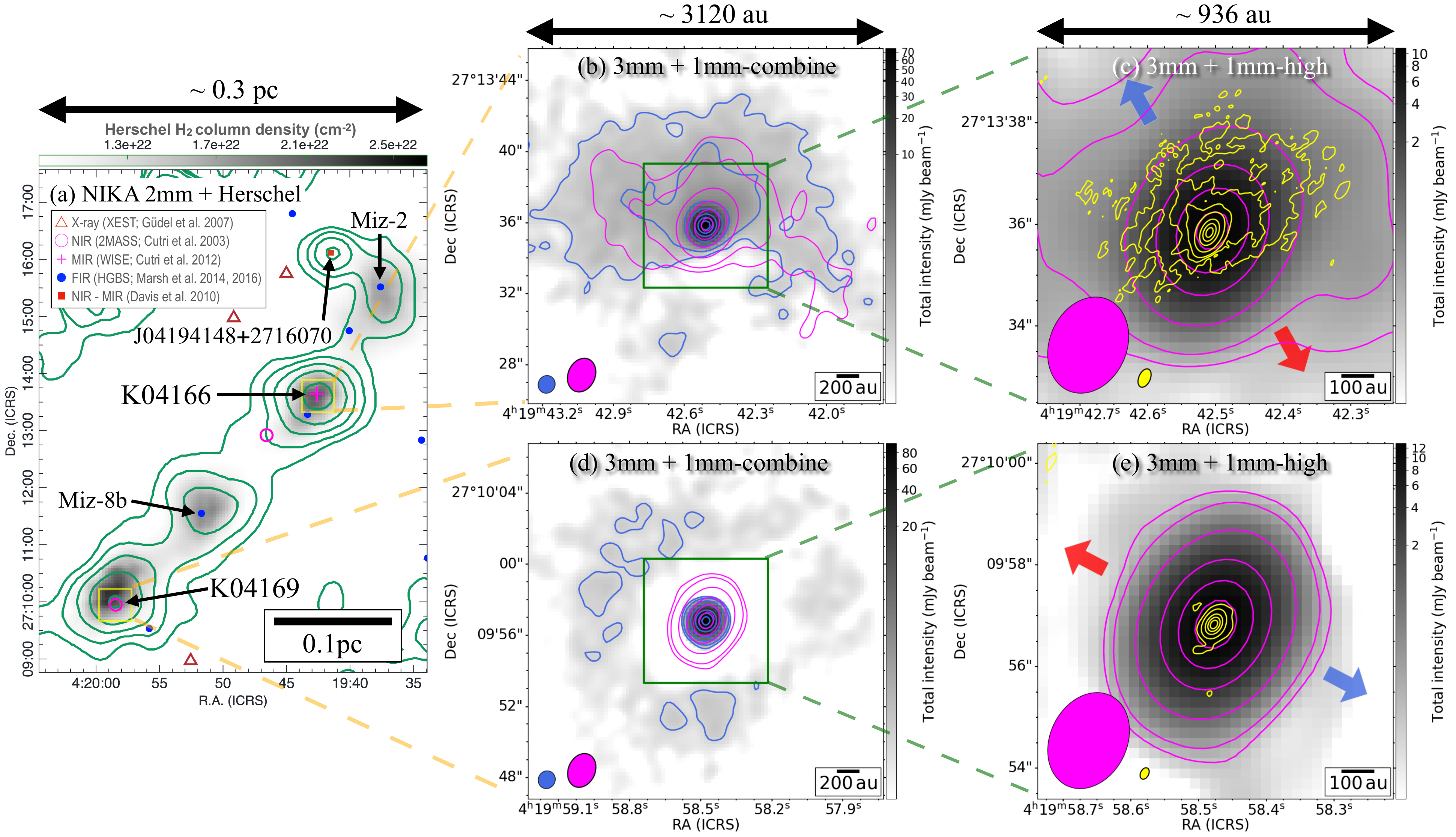}
     \caption{Overview of Stokes~$I$ dust continuum emission at different scales.
     Panel (a): Previous observations toward B213. Green contours represent the 2\,mm dust continuum emission derived from IRAM-30m/NIKA at the contour levels of $[2,5,10,20]\times0.068\,\mathrm{MJy\,sr^{-1}}$ \citep{Bracco2017A&A...604A..52B}. Background grayscale represents the Herschel $\mathrm{H_2}$ column density map. The symbols point to the source positions previously identified with X-ray \citep[red triangles;][]{Gudel_2007A&A...468..353G}, near-IR  \citep[NIR: magenta open circle;][]{Cutri_2003yCat.2246....0C}, mid-IR \citep[MIR: magenta cross;][]{Cutri_2012wise.rept....1C}, and FIR \citep[blue filled circles;][]{Marsh_2014MNRAS.439.3683M, Marsh_2016MNRAS.459..342M}, along with a T-Tauri source identified based on NIR-MIR \citep[the red filled square;][]{Davis_2010MNRAS.405..759D}.
     Panels (b) and (d): Zoom-in images of each source at $\sim$3000\,au scale. Background grayscale and blue contours represent the total intensity of the 1mm-combine continuum emission, with its contour levels of [3, 8, 30, 100, 300, 500, 700]$\times89\mathrm{\mu Jy\,beam^{-1}}$ in panel (b) and [3, 8, 30, 100, 300, 500, 700]$\times116\mathrm{\mu Jy\,beam^{-1}}$ in panel (d). The magenta contours represent the total intensity of the 3mm continuum, and the contour levels are [5, 10, 50, 200, 400, 600]$\times17\mathrm{\mu Jy\,beam^{-1}}$ in panel (b) and [5, 10, 50, 200, 400, 600]$\times18\mathrm{\mu Jy\,beam^{-1}}$ in panel (d).
     The blue and magenta ellipses at the bottom left correspond to the synthesized beam of the 1mm-combine and 3\,mm maps in both panels.
     Panels (c) and (e): Zoom-in images of panels (b) and (d) at $\sim$900\,au scale. Grayscale and the magenta contours represent the total intensity of the 3mm data, and the contour levels are the same as panels (b) and (d). Yellow contours represent the total intensity of the 1mm-high continuum, at [5, 10, 20, 100, 500, 1000, 1500, 2000, 2500, 3000]$\times33\mathrm{\mu Jy\,beam^{-1}}$ levels in panel (c) and [5, 100, 500, 1000, 1500, 2000, 2500, 3000]$\times28\mathrm{\mu Jy\,beam^{-1}}$ levels in panel (e). 
     Red and blue arrows represent directions of red- and blue-shifted $\mathrm{^{12}CO}$ outflow lobes with position angle of 32$\degr$ for K04166 \citep{Narayanan_2012MNRAS.425.2641N} and 64$\degr$ for K04169 \citep{Takakuwa_2018ApJ...865...51T}.
     Magenta and yellow ellipses at the bottom left correspond to the synthesized beam of the 3\,mm and 1mm-high maps in both panels.
     Root-mean-square (rms) values of the contours are listed in Table\,\ref{tab:para-continuum}.
     }
      \label{fig:NIKA-ALMA}
\end{figure*}

Figure~\ref{fig:NIKA-ALMA} presents a multiscale overview of the Stokes~$I$ dust continuum emission in B213 from cloud-to-core scales probed by single dish observations (NIKA 2\,mm map; \citealt{Bracco2017A&A...604A..52B}, panel\,a) to the core and disk scales sampled by our ALMA observations. In the data probing the core scales (NIKA map), similar dust thermal emission is associated with both protostars. However, at the smaller scale shown by our ALMA data, the emission morphology diverges between the two sources. The 1mm-combine (blue contours in panels\,b and d) and 3mm (magenta, panels\,b–e) maps reveal the dense, inner envelope ($\sim16\arcsec$ $\sim$ 2500\,au) around K04166, while in K04169 the emission is dominated by a compact component with hints of filtered extended emission (ring-like structure at a very low S/N). 

Finally, our highest resolution data (1mm-high; yellow contours, rightmost panels) highlight the dramatic difference between the two protostars: while K04166 is still associated with resolved emission, whose extent spans up to $\sim4\arcsec$ ($\sim$624\,au) at a 5$\sigma$ level, K04169 shows solely a compact component, with a size $\simlt 1\arcsec$ ($\sim$156\,au) at a 5$\sigma$ level. In Sect.~\ref{sec:contmodel}, we describe our modeling of the dust Stokes~$I$ emission from both sources to quantify the typical spatial extent of the emission, along with the respective contributions to the dust thermal emission from the disk component and from the envelope component.

\subsection{Disentangling disk and envelope contributions in the Stokes~$I$ dust continuum emission}
\label{sec:contmodel}

We quantified the spatial extent and origin of dust emission in both sources by modeling both the image and visibility data. In the image domain, we performed 2D Gaussian fits with CASA \texttt{imfit}, for each of the three datasets in each source and Table\,\ref{tab:imfit_results} lists the results. The 3mm data show unresolved emission for K04169 within the $\sim1.5 \arcsec$ beam, but K04166 remains resolved, indicating distinct underlying density structures. 
In the table, dust emission sizes derived from the 1mm-high dataset are $21 \times 15$\,au for K04166 and $30 \times 22$\,au for K04169: these sizes are consistent with the dust disk radii reported by the eDisk collaboration, using similar estimation methods \citep{Ohashi_2023ApJ...951....8O, Phuong_2025ApJ...992...18P, Han_2025ApJ...993..120H}. 

\begin{table*}
\centering
\scriptsize
\caption{Summary of 2D Gaussian model fitting of the Stokes~$I$ in the image plane.}
\label{tab:imfit_results}
\begin{tabular}{llllllllll}
\hline \hline
Source & Data & Peak position & Major axis \tablefootmark{a} & Minor axis \tablefootmark{a} & P.A. \tablefootmark{a} & Flux & Peak flux & Reference freq.  & Inc. \tablefootmark{b} \\
& & [$04^{\mathrm{h}}19^{m}..$, $+27^\circ..$] & (mas) & (mas) & (deg) & (mJy) & ($\text{mJy beam}^{-1}$)  & (GHz) & (deg) \\
\hline
K04166 & 3mm & {${\mathrm{..}}42^{\mathrm{s}}.504,\, \mathrm{..}13'35''.89$} & $729 \pm 13$ & $ 554 \pm 13$ & $127$ & $12.50 \pm 0.03$ & $10.96 \pm 0.02$ & 100.748 &  41\\
K04166 & 1mm-combine & {${\mathrm{..}}42^{\mathrm{s}}.509,\, \mathrm{..}13'35''.84$} & $402.0 \pm 5.6$ & $353.0 \pm 6.0$ & $119$ & $84.19 \pm  0.17$ & $73.27 \pm 0.09$ & 231.796 & 45 \\
K04166 & 1mm-high & {${\mathrm{..}}42^{\mathrm{s}}.509,\, \mathrm{..}13'35''.84$} & $136.6 \pm 0.5$ & $96.0 \pm  0.4$ & $125$ & $68.93 \pm  0.07$ & $50.76 \pm 0.03$ & 225.601 & 45 \\
\hline
K04169 & 3mm & {${\mathrm{..}}58^{\mathrm{s}}.474,\, {\mathrm{..}}09'56''.82$} & Unresolved & Unresolved & - & $ 13.13 \pm 0.03$ & $13.15 \pm 0.02$ & 100.748 & - \\
K04169 & 1mm-combine & {${\mathrm{..}}58^{\mathrm{s}}.478,\, {\mathrm{..}}09'56''.80$} & $169 \pm 11$ & $136 \pm 14 $ & $128$ & $98.30 \pm 0.21$ & $95.97 \pm 0.12$ & 231.796 &  36 \\
K04169 & 1mm-high & {${\mathrm{..}}58^{\mathrm{s}}.477,\, {\mathrm{..}}09'56''.83$} & $195.0 \pm 0.3$ & $143.0 \pm 0.2 $ & $139$ & $93.65 \pm 0.08$ & $54.132 \pm 0.030$ & 225.601 &  43 \\
\hline
\end{tabular}
\tablefoot{
Parameters obtained by minimizing an elliptical Gaussian model to the dust continuum emission maps, using the CASA task \texttt{imfit}.  
\tablefoottext{a}{Projected extent of the deconvolved size on the plane of the sky.}
\tablefoottext{b}{The inclination angles on the line-of-sight were estimated from the aspect ratio of major axis and minor axis best parameters, by assumption of a circular geometry for the disk. 0\degr means face-on, whereas 90\degr means edge-on.}
}
\end{table*}

\begin{figure*}[]
    \centering
    \includegraphics [width=17cm]{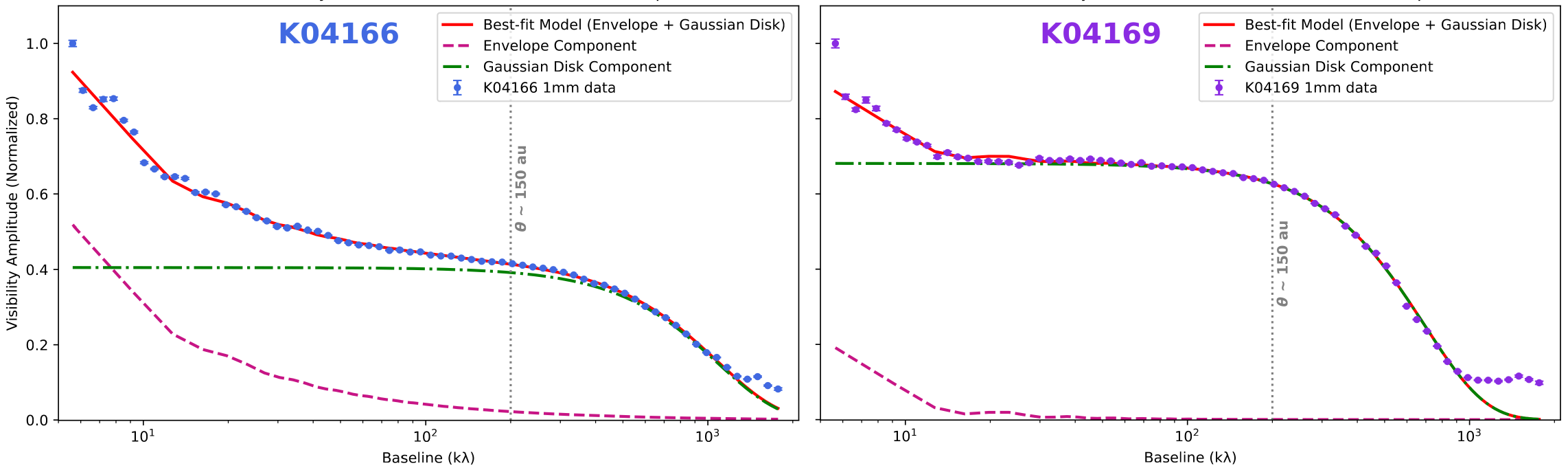}
     \caption{1.4\,mm dust continuum emission visibility amplitudes as a function of baseline length (circularly averaged in logarithmically spaced bins), for both K04166 (left) and K04169 (right). The visibility amplitudes were derived from 1mm-combine data and are normalized to the highest flux (at a baseline $5.6 \,\mathrm{k}\lambda$), for comparison purposes. A logarithmic scale is used for the uv-distance, as it encompasses several orders of magnitude, to enhance the visibility of all data points. The profiles of the two sources are shown, with associated error bars reflecting the dispersion of amplitudes in the bin and the best-fit models combining a Plummer envelope and a Gaussian disk are shown as the red curves on top of the data points.}
     \label{fig:uvfit_result_best}
\end{figure*}

However, modeling the spatial distribution of the flux density in the maps with a Gaussian model does not allow us to disentangle the emission due to the inner envelope from the one due to the disk, if the source is embedded. 
Establishing the characteristics of different structures in complex objects (e.g., embedded protostars), such as the size and flux of the disk while it is embedded in its parent envelope, requires a methodology that can discriminate between the objects on a given line of sight, as well as the respective contributions of both (e.g., disk and envelope emission). It has been demonstrated that the best method to measure disk sizes in embedded protostars is to directly model the data in the Fourier space, where the power at each spatial scale is directly measured in a unambiguous way; rather than in the emission maps, where all scales are contributing to the reconstructed emission \citep{Maury_2019A&A...621A..76M, Tung2024}.
Thus, we carried out a modeling of the visibility amplitudes for both sources, following \citet{Maury_2019A&A...621A..76M}. 
The uv-data used to perform the modeling include both the 7-m and 12-m arrays continuum visibility datasets obtained in Band\,6, listed as 1mm-combine data in Table\,\ref{obs_para}.
The analytical models were used alone or combined, then minimized to reproduce the azimuthally averaged amplitude profiles of each source.
Table \ref{tab:uvmodel_all_am} summarizes the results from modeling the visibility amplitudes with Plummer envelopes, both including and not including Gaussian disks and uniform disks components.

\begin{table*}
\centering
\scriptsize
\caption{Disk and envelope contributions: models of the 1.4\,mm Stokes~$I$ visibility profiles.}
\label{tab:uvmodel_all_am}
\begin{tabular}{llccccccccc}
\hline\hline
Source & Model \tablefootmark{a}  & Envelope flux & Envelope $p+q$ & Envelope $R_{in}$ & Envelope $R_{out}$ & Disk flux & Disk radius \tablefootmark{b} & $\chi^2$ \tablefootmark{c} & Best model? \tablefootmark{d} \\
& & (mJy) &  & (arcsec) & (arcsec) & (mJy) & (milliarcsec - au) & & \\
\hline
K04166 & P &   247 (41) & 2.5 (0.2) & 0.001 (0.007) & 6 (2) & - & - & 1298 & no \\
K04166 & G & - & - & - & - & 74 (1) & 109 (7) - 17 & 72 & no  \\
K04166 & D & - & - & - & - & 73 (1) & 85 (5) - 13 & 82 & no \\
K04166 & PG & 153 (2) & 2.2 (0.1) & 0.01 (0.05) & 15 (1) & 67 (2) & 100 (2) - 16 & 2 & yes \\
K04166 & PD & 150 (3) & 2.3 (0.1) & 0.01 (0.03) & 15 (1) & 65 (3) & 80 (2) - 12& 5 & no\\
\hline
\hline
K04169 & P &   293 (20) & 2.5 (0.2) & 0.001 (0.008) & 6 (3) & - & - & 1087 & no \\
K04169 & G & - & - & - & - & 99 (1) & 158 (3) - 25 & 7 & no \\
K04169 & D & - & - & - & - & 97 (1) & 121 (2) - 19 & 15 & no \\
K04169 & PG & 126 (5) & 1.3 (0.5) & 0.01 (7) & 13 (3) & 98 (1) & 157 (3) - 24 & 5.7 & yes \\
K04169 & PD & 115 (4) & 2.8 (0.3) & 0.01 (0.6) & 15 (2) & 86 (1) & 120 (2) - 19 & 9 & no \\
\hline
\end{tabular}
\tablefoot{ Best-fit parameters for the four models of disk and/or envelope explored to reproduce the emission of K04166 and K04169 at 1.4\,mm.
\tablefoottext{a}{P, G, D, PD, and PG stand for Plummer-envelope-only, Gaussian-disk-only, uniform-disk-only, Plummer-envelope+uniform-disk, and Plummer-envelope+Gaussian-disk models, respectively (see \citealt{Maury_2019A&A...621A..76M} for more description of the modeling of visibilities).}  
\tablefoottext{b} he disk radius reported is either the Gaussian FWHM for the Gaussian disk model, or the radius from the uniform disk model. The radii in astronomical units are given, using a distance of 156\,pc. 
\tablefoottext{c}{$\chi^2$ corresponds to the $\chi^2$ per degree of freedom (reduced $\chi^2$).}
\tablefoottext{d}{Reports whether the model is the best at reproducing the data according to the F-test.}
}
\end{table*}

The best models obtained for both sources are shown in Fig.~\ref{fig:uvfit_result_best}.
We stress that while K04166 requires a disk and a significant contribution from a dense envelope to be modeled satisfactorily (models with a Plummer envelope and a disk show reduced $\chi^2$ 2 to 3 orders of magnitude better than models including only a disk or an envelope, see Table \ref{tab:uvmodel_all_am}), K04169 is satisfactorily described by models including only a disk component although models including a diffuse envelope contribution slightly improve the reduced $\chi^2$ (from 7.1 to 5.7). 
Our modeling estimates the disk radii to be 16\,au for K04166 and 24\,au for K04169. Most ($\sim 70\%$) of the flux in the 1mm-combined data comes from the disk component in K04169, while only $\simlt 40\%$ of the total flux is due to the disk contribution in K04166, barely resolved with our data (see Fig.~\ref{fig:uvfit_result_best}). We note the presence of a $\sim 5$\,mJy unresolved component at the longest baselines ($> 1000 \,k\lambda$) in K04169, which was not modeled by our combination of envelope and disk components. This component could be tracing an optically thick inner disk or a disk substructure. Similarly, in K04166, our simple analytical disk model does not reproduce a feature seen as a flattening at small disk radii < 15\,au (>1000\,$\mathrm{k\lambda}$), most likely due to a constant brightness region where dust emission becomes optically thick.

Finally, we note that these models cannot faithfully characterize the properties of the outer envelope(s) at scales $\simgt 3000$ au, as the data are intrinsically not probing any scale beyond $\sim12 \arcsec$, despite the use of the ACA data in the modeling. Nevertheless, no peaked density profile suggestive of the presence of an envelope was found around K04169 at the scales (i.e., 20-3000\,au) probed by our data.
Moreover, we find that the disk radius in K04166 ($\simeq 16$\,au) is tentatively smaller than the K04169 one ($\simeq 24$\,au), and smaller than the extent of the continuum emission probed with imfit, in the 1mm-combine map that was built from the same visibilities as those modeled here.

\subsection{Polarized dust emission in K04166}

\begin{figure*}[h!]
    \centering
    \includegraphics [width=17.5cm]{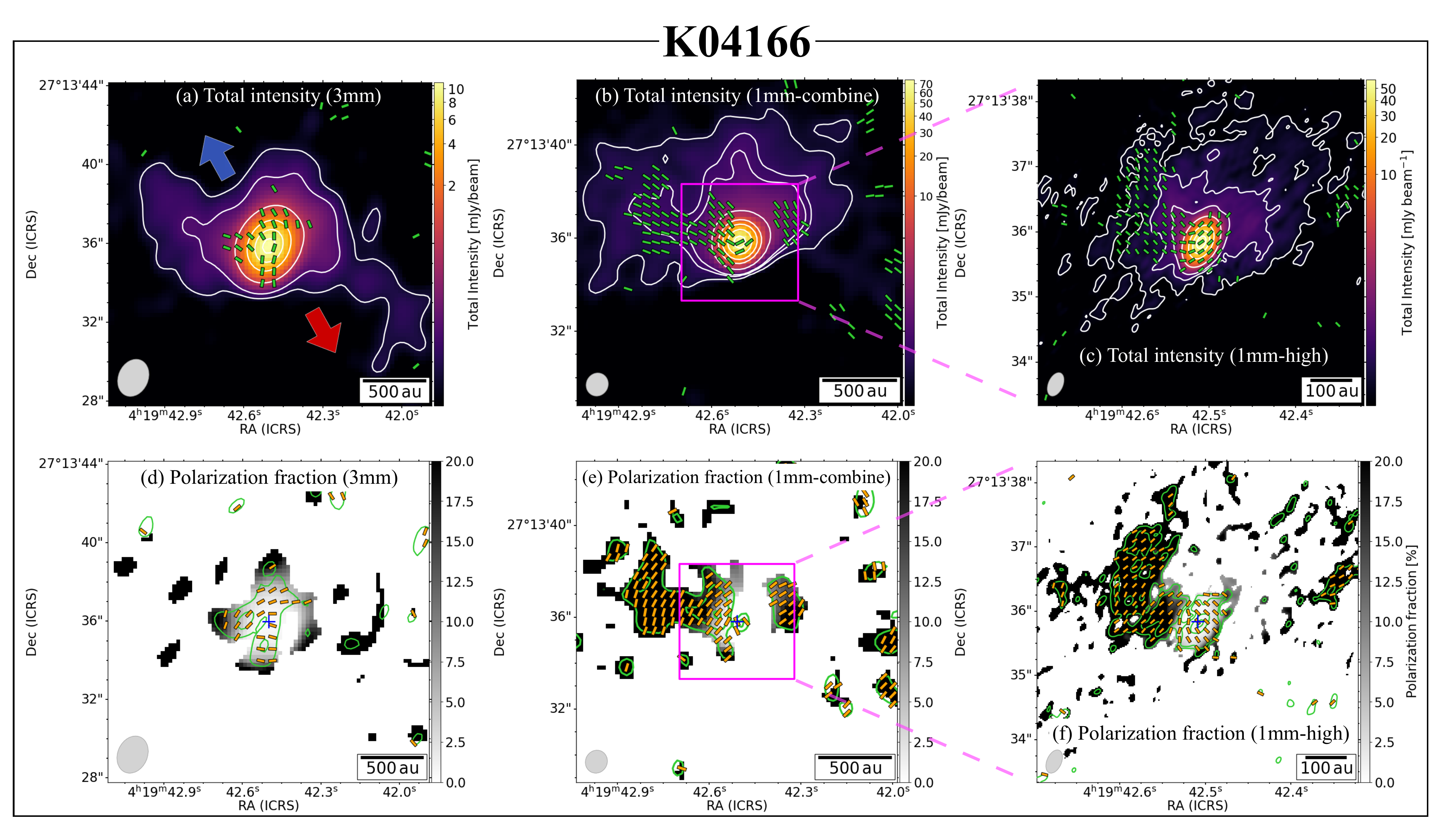}
     \caption{Polarization properties for K04166; 3\,mm at $\sim$230\,au resolution (panels a and d), 1mm-combine at the $\sim$156\,au resolution (panels b and e), and 1mm-high at $\sim$36\,au resolution (panels c and f). 
     Top row:  Total intensity at [5, 10, 50, 200, 400, 600]$\times17\,\mathrm{\mu Jy\,beam}^{-1}$, [5, 8, 10, 30, 100, 200, 400, 700]$\times89\,\mathrm{\mu Jy\,beam}^{-1}$, and [5, 10, 20, 100, 500, 1000, 1500, 2000, 2500, 3000]$\times33\,\mathrm{\mu Jy\,beam}^{-1}$ in panels a, b, and c, respectively, shown as the color scale and white contours. The green segments denote the B-vectors. The 1$\sigma$ for the white contours are listed in Table\,\ref{tab:para-continuum}. 
     The blue and red arrows in panel a indicate the direction of the blue- and red-shifted outflow lobes. 
     The magenta square in panel b corresponds to the map size in panels c and f, respectively. The map size is the same between panels a, b, d, and e.
     Bottom row:  Polarization fraction and polarized intensity, respectively, shown as the grayscale and green contours. The contour levels are [3, 5]$\times16\,\mathrm{\mu Jy\,beam}^{-1}$, [3, 5, 8]$\times47\mathrm{\mu Jy\,beam}^{-1}$, and [3, 5, 15, 30]$\times18\,\mathrm{\mu Jy\,beam}^{-1}$ in panels d, e, and f, respectively. The blue crosses point to the peak of total intensity of the same data as in the top row. The orange segments are the polarization vectors, perpendicular to the B-vectors in the top panel.
     The synthesized beam is presented at the bottom left side of each panel. We note that all segments in this paper are shown at fixed lengths, which do not relate to the polarization fraction value, as the following figures in this paper also present the segments with a fixed length.}
      \label{result_K66_polari}
\end{figure*}

In Fig.~\ref{result_K66_polari}, we present the polarized dust emission observed toward K04166 across our three datasets, spanning large (left panel) to small scales (right panel). At large scales (leftmost panels), the 3\,mm map appears to exhibit an elongated structure in Stokes~$I$ from northeast to southwest (panel a); whereas the polarized intensity exhibits a double-peaked structure, offset from the Stokes~$I$ peak,  suggesting a decrease in polarized intensity along the southeastern equatorial plane (panel d).

In the 3\,mm map, the polarization vectors are neither aligned with the outflow axis nor the disk major axis. Instead, the inferred B-vectors (obtained by rotating the polarization vectors by $90\degr$) display roughly an hourglass-shaped morphology \citep[e.g.,][]{Girart_2006Sci...313..812G, Frau_2011A&A...535A..44F}.

The 1mm-combine and 1mm-high maps reveal extended polarized emission elongated toward the northeast from the total intensity peak. Here, B-vectors almost align with the outflow axis in the outer regions, while (most significantly in the 1mm-high map), they are perpendicular to the disk minor axis in the inner regions. The histogram of polarization position angles (Fig.~\ref{fig:PA_histogram}, left panel), fitted with a two-component Gaussian, shows peaks separated by 90\degr, at $43.8\degr$ and $131.0\degr$, indicating the inner and outer regions, respectively.

In the bottom row of Fig.~\ref{result_K66_polari}, the polarization fraction in the 3\,mm maps is below 20\% for regions above $3\sigma$ of the polarized intensity. The 1mm-combine and 1mm-high maps show polarization fractions ranging from $\sim$5 to 20\% in the outer region (second Gaussian component in the histogram), and below $\sim$3\% in the inner region (the first component). 

Additionally, in all bottom-row maps, the peak of polarized intensity is offset from the peak of total intensity. In the 3\,mm map, the two polarized peaks are offset by $\sim1.1\arcsec\pm0.1\arcsec$ and $\sim1.2\arcsec\pm 0.1\arcsec$ from the total intensity peak, and in the 1mm-combine map, the shift is $0.20\arcsec\pm 0.09\arcsec$. 
Finally, the 1mm-high map shows its peak polarized intensity is offset from the Stokes~$I$ peak by $0.040\arcsec \pm 0.004\arcsec$, along the outflow axis \footnote{Note: the errors stem from the ALMA relative positional accuracy between the positions of two sources or features in a map, as given by $\sim0.5\times \mathrm{beam_{FWHM} / SNR}$ \citep{position_accuracy}.}. An interpretation of the origin of the polarized emission in K04166 is provided in the discussion in Sect.~\ref{sub:discussion-K66}.

\subsection{Polarized dust emission in K04169}\label{ResultsK04166}

\begin{figure*}[h!]
    \centering
    \includegraphics [width=17.5cm]{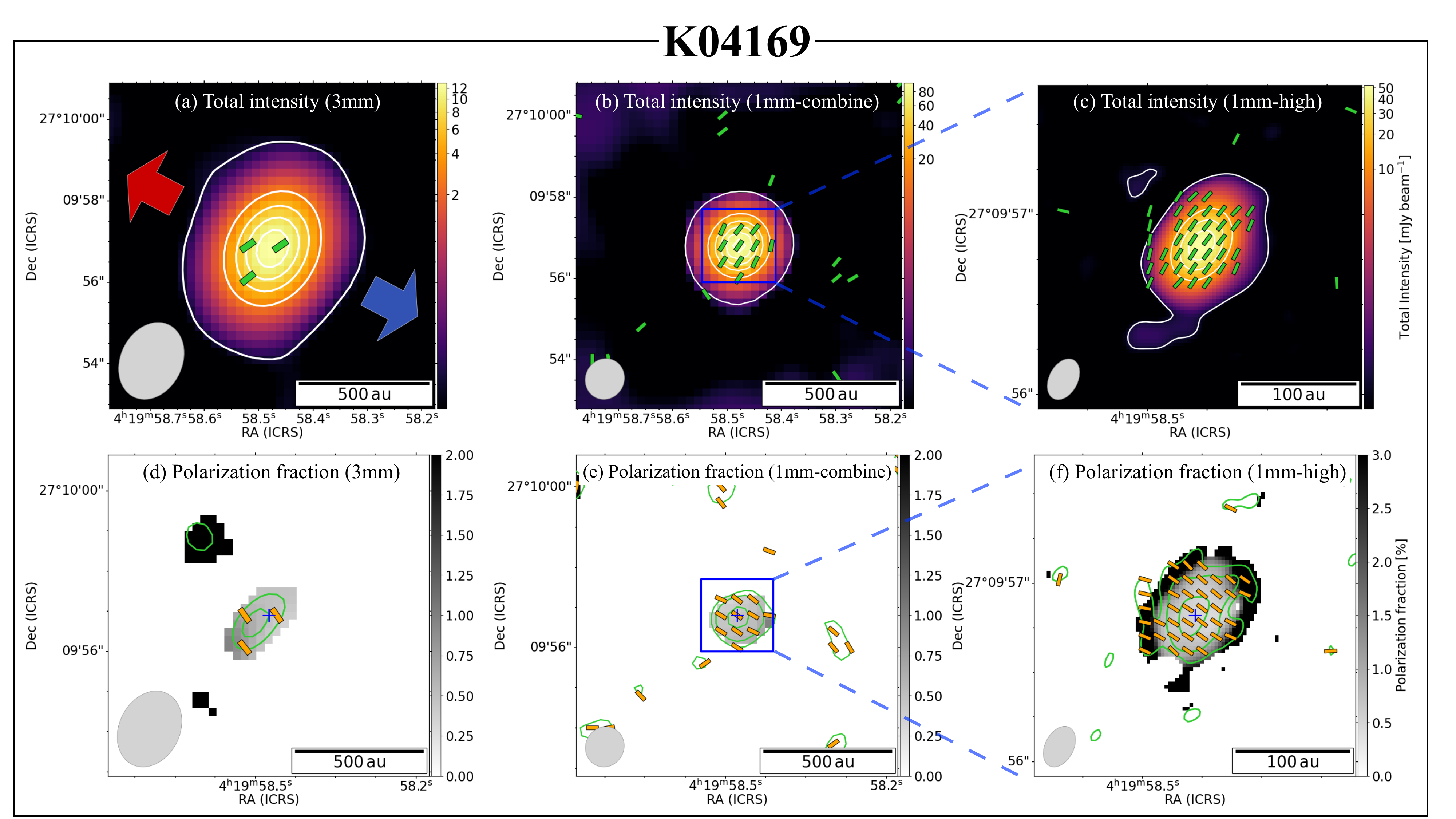}
     \caption{Polarization properties for K04169; 3\,mm (panels a and d), 1mm-combine (panels b and e), and 1mm-high (panels c and f).
     Top row:  Total intensity at [5,155,305,455,605]$\times18\,\mathrm{\mu Jy\,beam}^{-1}$, [5, 155, 305, 455, 605, 755]$\times116\,\mathrm{\mu Jy\,beam}^{-1}$, and [5, 505, 1005, 1505, 2005, 2505, 3005]$\times28\,\mathrm{\mu Jy\,beam}^{-1}$ in panels a, b, and c, respectively, shown as the color scale and white contours. The green segments denote the B-vectors. 1$\sigma$ for the white contours are listed in Table\,\ref{tab:para-continuum}. The
     1$\sigma$ for the white contours are listed in Table\,\ref{tab:para-continuum}. 
     The blue and red arrows in panel a indicate the direction of the blue- and red-shifted outflow lobes. 
     The magenta square in panel b corresponds to the map size in panels c and f, respectively. The map size is the same between panels a, b, d, and e.
     Bottom row: the grayscale and green contours represent the polarization fraction and polarized intensity, respectively. The contour levels are [3, 4]$\times15.8\,\mathrm{\mu Jy\,beam}^{-1}$, [3, 5, 10]$\times35.0\,\mathrm{\mu Jy\,beam}^{-1}$, and [3, 5, 10]$\times20.1\,\mathrm{\mu Jy\,beam}^{-1}$ in panels d, e, and f, respectively. The blue crosses point to the peak of total intensity of the same data as in the top row. The orange segments are the polarization vectors, perpendicular to the B-vectors in the top panel.
     The synthesized beam is presented at the bottom left side of each panel.}
      \label{result_K69_polari}
\end{figure*}

In Fig.~\ref{result_K69_polari}, we can see that all the maps exhibit more compact structures than K04166 in the total intensity (the top row), and both the 3\,mm and 1mm-combine polarized emission are basically unresolved, peaking roughly at the disk position.
In contrast, the 1mm-high polarized emission is extended, with a size of $\sim0.6\arcsec$ at $3\sigma$, and its peak is within the disk, as shown in the bottom row. This suggests that the polarized emission in the three maps comes from the disk. The polarization fraction in the three maps is also below 2\%. 

\begin{figure*}[h!]
    \centering
    \includegraphics [width=17cm]{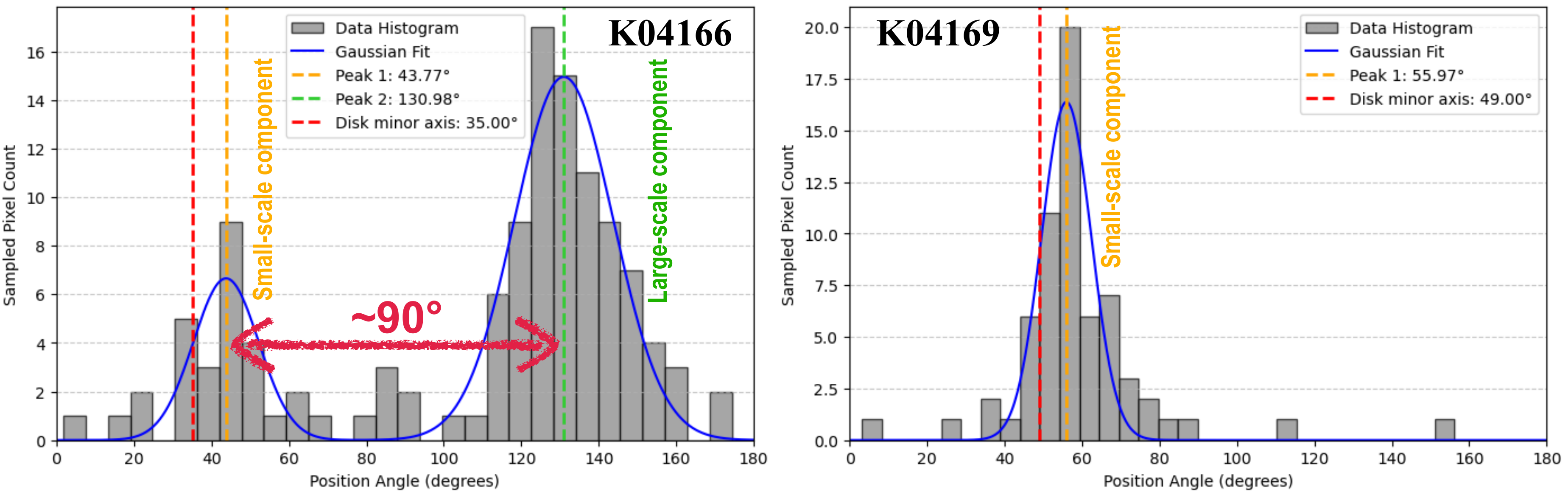}
   \caption{Histogram of position angles of the polarization vector around the disks, i.e., from the 1mm-high data.
   Left: Histogram for K04166 with a two-component Gaussian fitting denoted by the blue solid line. The gray-shaded bins were derived within a circle with a radius of 1.5\arcsec from the peak position in the total intensity, covering the extended polarized emission around the source.
   The dashed yellow and green vertical lines point to the position angles at the fitting peaks.
   The dashed red vertical line corresponds to the disk's minor axis in the continuum data (Table\,\ref{tab:imfit_results}).
   Right:  Histogram for K04169 with a one-component Gaussian fitting denoted by the blue solid line. The gray-shaded bins were derived within a circle with a radius of 0.5\arcsec from the peak position in the total intensity covering the compact structure in polarized intensity associated with K04169.
   The dashed yellow and red vertical lines are the same as those in the left panel.
   }
      \label{fig:PA_histogram}
\end{figure*}

The polarization angle distribution in the 1mm-high map peaks at a position angle of $56\pm 1$\degr \ and it shows a narrow distribution, almost parallel to the disk minor axis (see right panel of Fig.~\ref{fig:PA_histogram}).
The polarization angles from the 1mm-combine and 3\,mm polarization maps, at the polarized peak intensity, are measured to be $55.41\degr \pm 1.15\degr$ and $35.38\degr \pm 4.78\degr$, respectively. Within the uncertainties, the position angle of the polarization vector in the 1mm-combine maps is roughly in agreement with the 1mm-high angle distribution; however, the 3\,mm position angle stands out.
\footnote{Note: the position angle errors are estimated as $28.65\degr$/SNR, with S/N as the signal-to-noise ratio of the polarized intensity, from \citet{Serkowski_1958AcA.....8..135S, Naghizadeh-Khouei_1993A&A...274..968N}.}

Additionally, all bottom-row maps show a positional offset between the peaks of total and polarized intensity, oriented roughly perpendicular to the outflow axis, although the offset is within the disk size. All the offsets are significant, as measured to be $0.04\arcsec \pm 0.005\arcsec$ for the 1mm-high map, $0.50\arcsec \pm 0.16\arcsec$ for the 3mm map, and $0.06\arcsec \pm 0.02\arcsec$ for the 1mm-combine map. An interpretation of the origin of the polarized emission in K04169 is provided in the discussion in Sect.~\ref{sub:discussion-K69}. 

\section{Discussion}
\label{sec_4}

\subsection{Optical depths of the dust thermal emission}
\label{sect_brightness_temp}
We calculated the brightness temperatures at the peak of the dust continuum emission from the 1mm-high maps, following 
 \begin{equation}
      T_B = 1.222 \times 10^3 \, 
      \left[ \frac{I}{\rm mJy \, beam^{-1}} \right] \,
      \left[ \frac{\nu}{\rm GHz} \right] ^{-2} \,
      \left[ \frac{\theta_\mathrm{maj}}{''} \right]^{-1} \,
      \left[ \frac{\theta_\mathrm{min}}{''} \right]^{-1} \,
 ,\end{equation} 
where $T_B$ is the brightness temperature, $I$ is the peak intensity, $\nu$ is the frequency (218.4\,GHz here), and $\theta_\mathrm{maj}$ and $\theta_\mathrm{min}$ are the half-power beam widths along the major and minor axes, respectively. We find brightness temperatures at the peak of the dust continuum emission of 30.7\,K for K04166 and 33.8\,K for K04169. However, the disks are only partially resolved, since their deconvolved FWHM (see Table~\ref{tab:imfit_results}) are slightly smaller than the 1mm-high beam. That means that the estimated brightness temperature values are beam diluted. For a Gaussian, the filling factor is defined as 
$f \, = \, \theta_\mathrm{src}^2/(\theta_\mathrm{src}^2+\theta_\mathrm{bm}^2)$, 
where 
$\theta_\mathrm{src}$ and $\theta_\mathrm{bm}$ 
are the geometrical FWHM average of the disks and of the beam, respectively. After correcting for filling factor, we found the brightness temperature of 90\,K for K04166 and 47\,K for K04169. 
These are lower than peak brightness temperatures obtained at higher angular resolution, 133\,K and 64\,K, respectively \citep{Ohashi_2023ApJ...951....8O}.
Furthermore, given their bolometric luminosities, the expected dust temperatures at their disk radius (Table~\ref{tab:uvmodel_all_am}) are $\sim$75\,K and 80\,K for K04166 and K04169, respectively. This indicates that the dust emission is fully optically thick for K04166 and moderately optically thick for K04169. According to \citet{Yang_2017MNRAS.472..373Y}, a dust optical depth of $\sim$1--2 is sufficient to produce 1--2$\%$ levels of polarized emission from self-scattering. Thus, at the beam scales characterizing our data, the polarized emission should be interpreted on the basis of this hypothesis.

\subsection{Nature of the polarized emission in K04166}
\label{sub:discussion-K66}

\begin{figure}
  \resizebox{\hsize}{!}{\includegraphics{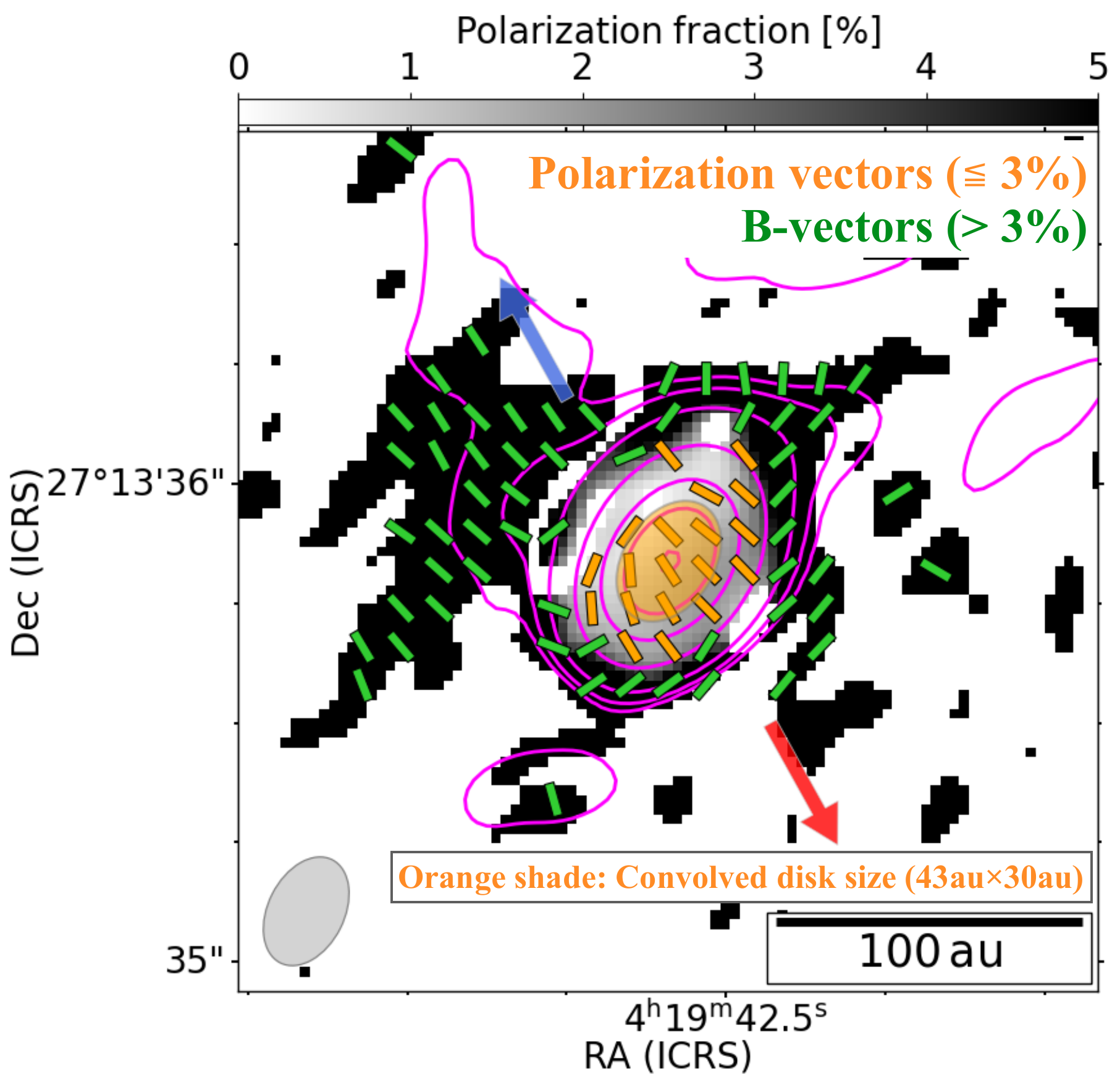}}
      \caption{Central region of K04166 obtained from the 1mm-high data with Briggs weighting. The background grayscale represents the polarization fraction, and the magenta contour represents the total intensity at [10, 20, 50, 200, 600, 1200, 2000]$\times25\,\mathrm{\mu Jy\,beam}^{-1}$ levels. The orange segments correspond to the polarization vectors within the area where the fraction is lower than 3\,\%, whereas the green segments correspond to the 90$\degr$ flipped polarization vectors, B-vectors, within the area where the fraction is larger than 3\,\%.
      The orange-shaded ellipse corresponds to the convolved size obtained with 2D Gaussian fitting. The contour at 200$\times25\,\mathrm{\mu Jy\,beam^{-1}}$ levels is comparable to the transition area of 3\%, which has the projected size of $\sim$0.51\arcsec$\times$0.36\arcsec (80\,au$\times$56\,au). The blue and red arrows are the outflow direction, and the gray ellipse corresponds to the beam.}
      \label{fig:K66_dis_1}
\end{figure}

Our observations of K04166 reveal tentative evidence for a scale-dependent transition in the dominant dust polarization mechanism. The 3mm maps indicate an extended component of the polarization, beyond 500\,au scales, with a tentative hourglass-shaped magnetic field (see panel a of  Fig.~\ref{result_K66_polari}): this polarized emission has relatively high polarization fraction values ($\gtrsim 5$\%),
suggesting that in these regions, polarized dust emission traces magnetically aligned grains. Our observations are consistent with previous observations tracing magnetic field lines pinched by gravitational infall in protostellar envelopes, commonly referred to as the ``hourglass'' magnetic field morphology \citep{LeGouellec_2019ApJ...885..106L, Huang_2024ApJ...963L..31H}. 

In contrast, the 1mm-high and 1mm-combine maps (panels b and c of Fig.~\ref{result_K66_polari}) still trace this large-scale component, but also reveal the polarization vectors differing from the large-scale component by 90\degr in the inner regions (the left panel of Fig.~\ref{fig:PA_histogram}). Previous studies have shown that self-scattering-induced polarization is characterized by the polarization vectors parallel to the disk minor axis and low polarization fractions, supporting our interpretation \citep{Kataoka_2015ApJ...809...78K, Huang_2024ApJ...963L..31H, Liu_2024ApJ...963..104L}. Thus, our observations could be interpreted as a transition, in K04166, from polarization due to self-scattering at disk scales to magnetically aligned grains at envelope scales. 

To distinguish the possible different origin of the polarized emission, we adopt a 3\% polarization fraction threshold on the highest resolution 1mm-high map, to either (i) rotate by 90\degr the polarization angles if the polarization fraction is larger than this threshold or (ii) keep the polarization angle below this threshold, which occurs around or within the disk. In this latter case, the polarization angle is, on average, parallel to the disk's minor axis.  This is a strong indication of self-scattering  \citep[e.g.,][]{Kataoka_2015ApJ...809...78K, Brunngraber2021A&A}. In Fig.~\ref{fig:K66_dis_1}, we see the spatial distribution obtained between these two regimes assuming this ad hoc, simple threshold. We note that at the boundary where the polarization vectors switch to B vectors, the polarization fraction drops to nearly zero on the northern side of the disk. This coincides with a steady increase in total intensity, which could be interpreted as a result of cancellation between the self-scattering polarization component and the magnetic alignment polarization component. 

Additionally, the area in which the polarization fraction stays low exceeds that of the disk's extent (orange ellipse in Fig.~\ref{fig:K66_dis_1}), which could either be due to a beam effect spreading the polarized signal from the disk, or to some polarized emission due to self-scattering produced in the inner envelope. 
However, we stress that the precise location of the transition between polarization mechanisms depends on how both mechanisms are discriminated, which is itself sensitive to model assumptions. 
In our data, the 3mm map, with its tentative hourglass morphology and large polarization fraction, strongly suggests the presence of elongated grains because only elongated grains would be able to align with the local magnetic field.
This interpretation leads to two possibilities for the observed transition between the envelope and the disk, where the polarization angle appears to be parallel to the disk minor axis, but with two distinct regions, one with low polarization fraction, $\lesssim 2$\%, and the other with high values,$\gtrsim 4$\%  (orange and green segments, respectively, in Fig.~\ref{fig:K66_dis_1}).
First, the observed pattern may trace a toroidal magnetic field in the envelope, close to the disk’s outer radius. 
If this is the case, the large polarization fraction at the large envelope densities expected at disk's scales could be explained by recent theoretical studies, where mechanical and enhanced radiative torque alignment may be important  \citep{Hoang2022, Giang2024, Giang2025}.
Alternatively, the emission results from self-scattering in both cases, indicating that large grains exist outside the disk, in the inner envelope.
Theoretical studies suggest that self-scattering by spherical grains can produce polarization fractions of up to a few percent \citep{Kataoka_2015ApJ...809...78K}, but models with nonspherical grains allow for larger values \citep{Kirchschlager_2020A&A...638A.116K} of the polarization fraction. 

If the detection of self-scattering were eventually confirmed in the envelope of K04166, this could provide an additional strong observational evidence for the presence of significantly grown dust grains in the innermost regions of protostellar envelopes, which would be in agreement with the small emissivity indices of dust sometimes measured at millimeter wavelengths in these environments  \citep{Galametz2019, Cacciapuoti2025}. However, we stress that the typical grain size needed to produce the self-scattering signal is subject to large uncertainties, as self-scattering depends not solely on grain size but also on the dust structure and optical properties, as discussed above. These findings may motivate future theoretical work to refine predictions regarding the polarization mechanisms at play when physical conditions allow both magnetically aligned grains and self-scattering to produce polarized emission.
These findings could also support the enhanced presence of large, millimeter-sizes, grains lifted from the disk. For example, \citet{Cacciapuoti_2024} reported an anti-correlation between the dust opacity index ($\beta$) in the inner envelope and the mass-loss rate of jets and outflows. This suggests that grains with large sizes detected in envelopes could come from dust uplifted from the disk via outflows. This scenario is supported by both analytical and MHD models \citep{Wong_2016_10.1093/pasj/psw066, Lebreuilly_2020A&A...641A.112L, Tsukamoto_2021, Bhandare2024} and it could also explain the observed polarization patterns in the transition region.
Future observations with ALMA at different frequencies (Bands 1 and 10), at comparable resolution, will be crucial to disentangling the contributions of self-scattering and magnetic alignment, while also linking them to measurements of the dust $\beta$. Such multifrequency, high-angular resolution data will provide new constraints on the processes responsible for dust evolution, dust transport, and the robustness of millimeter polarized dust emission to trace the magnetic field structure down to very small scales in protostellar systems.

Finally, Fig.~\ref{fig:outflow_polari} shows the gas tracing the outflows for both sources from our data. The left panel of Fig.~7 exhibits that, in the high-resolution data (1mm-high and 1mm-combine), the polarized emission is highly asymmetric, with the detection of B-vectors seen mostly toward the blue-shifted outflow lobe. In this region, the polarization is enhanced along outflow cavity walls; in addition, its magnetic field direction at these 500--1000 au scales has a similar orientation as the large-scale B-field threading the B213 filament \citep{Palmeirim_2013A&A...550A..38P}. Such enhanced polarized emission has been reported before in high angular resolution observations of protostars; for instance, in B335 and Serpens \citep{Maury2018, LeGouellec_2019ApJ...885..106L}. It is also occasionally associated with the locations where molecular tracers indicate high ionization fractions. Thus, it has been interpreted as enhanced dust alignment efficiency along the magnetic field lines in these regions, which receive larger fluxes of ultraviolet and optical photons able to align the small grains, owing to the opening of the outflow cavity where photons from the central protostar can propagate further than along the high-density sightlines through the equatorial plane \citep{LeGouellec_2023A&A...675A.133L}.

\subsection{Nature of the polarized emission in K04169}
\label{sub:discussion-K69}

In Sect.~\ref{ResultsK04166}, we described how K04169 exhibits compact polarized emission, low polarization fractions, and the polarization vectors approximately parallel to the disk minor axis.  Furthermore, the polarized intensity peak in K04169 is offset from the peak of the dust thermal emission (Stokes~$I$).
These features suggest that the polarization at 1\,mm is dominated by self-scattering of dust grains from the disk \citep{Kataoka_2015ApJ...809...78K}. We note that the polarization fractions at radii $> 30$ au show high values in the map (see Fig.~\ref{result_K69_polari}f), whereas the polarized intensity has an extent larger (deconvolved size of 64\,au, see Table~\ref{tab:imfit_pol_app}) than the thermal dust continuum emission extent (deconvolved size 30\,au). This could stem from a true difference in the spatial extents of the polarized and Stokes I emission or from intrinsically high polarization fractions at such scales. 

The 3\,mm polarization from K04169, on the other hand, exhibits a weak compact detection, and the peak of polarized intensity is significantly offset from the peak of Stokes~$I$ from the high-resolution 1\,mm data. Moreover, the polarization angle is not consistent with the polarization angle measured in the 1mm maps: the differences are significant between both the 3\,mm and 1mm-high maps ($35.4\pm4.8$\degr and $56.0\pm1.0$\degr, respectively). Despite a low signal-to-noise ratio (S/N of 6 at the peak for the 3\,mm map), these may indicate the presence of another component contributing to the detected polarized emission at 3\,mm, either magnetically aligned grains from the surrounding material, or optical depth effects affecting the disk polarization differently at 1\,mm and 3\,mm.
In addition, higher sensitivity and angular-resolution observations are necessary to characterize the origin of the 3\,mm polarization signal in K04169.
In such a case, the orientation of the polarization vectors suggests either magnetically aligned dust grains along a toroidal B-field, or polarization due to self-scattering which would tentatively suggest some dust growth in the inner envelope where high-density materials surround the disk.
The detection of polarized emission due to self-scattering in disks has been reported, either in Class\,II YSO \citep[e.g.,][]{Bacciotti_2018ApJ...865L..12B, Dent2019MNRAS.482L..29D}, or in embedded protostars \citep{Harris2018, Sadavoy2019, CFLee2021, Huang_2024ApJ...963L..31H}. It has been interpreted as evidence of the presence of large grains in these young disks. Following computations from the literature, our detection of self-scattering at the disk scales in K04169 could suggest dust grains as large as hundreds of micrometers. However, as discussed before, large uncertainties remain when attempting to compute a maximum grain size from the detection of millimeter polarized emission associated with self-scattering.

Interestingly, previous 850~$\mu$m JCMT observations (probing $\sim$0.01\,pc scales) display a polarization angle of $59\pm7$\degr  \citep[at the peak of the total intensity:][]{Chakali_2021ApJ...912L..27E}, which is similar to the ALMA 1\,mm polarization angles, $\simeq56$\degr. Furthermore, the polarization fraction from JCMT data at this position is $0.8\pm0.2\%$, consistent with the ALMA 1mm value ($\sim$0.5\%). 
This indicates that the JCMT polarization signal is likely dominated by self-scattering from the compact disk mapped by ALMA and, thus, it might not trace magnetic fields toward K04169. This is also supported by the fact that the orientation of polarization vectors from the single-dish is almost perpendicular to the B-field lines at the filament scale \citep[$\sim$0.1\,pc;][]{Palmeirim_2013A&A...550A..38P}. Thus, we stress that the magnetic-field strength estimate proposed from the JCMT data toward K04169 might have to be revised.

The right panel of Fig.~\ref{fig:outflow_polari} shows the $\mathrm{^{12}CO}$ line emission tracing the outflow associated with K04169. We note that the $\mathrm{^{12}CO}$ outflow detected in our high angular resolution observations of K04169 shows a different orientation than the large-scale outflow reported by \citet{Narayanan_2012MNRAS.425.2641N} ($\sim$30\degr), and more in agreement with the orientation reported by \citet{Tokuda_2020ApJ...899...10T} with the ACA observations sampling the 1000\,au scales ($\sim$64\degr). 
We did not further analyze the molecular line emission; however, the dynamical time of the outflow detected in our data (a few $10^3$\,yr) is estimated to be shorter than that of the outflow identified by \citet{Narayanan_2012MNRAS.425.2641N} ($\sim7\times10^4$\,yr). 
These different outflow position angles and different dynamical times may indicate precession, suggesting a possible link between polarization orientation, dust evolution, and toroidal B-field structure.

\subsection{Evolution: Comparing the twin protostars}

\begin{figure*}[h!]
   \centering
   \includegraphics[width=\hsize]{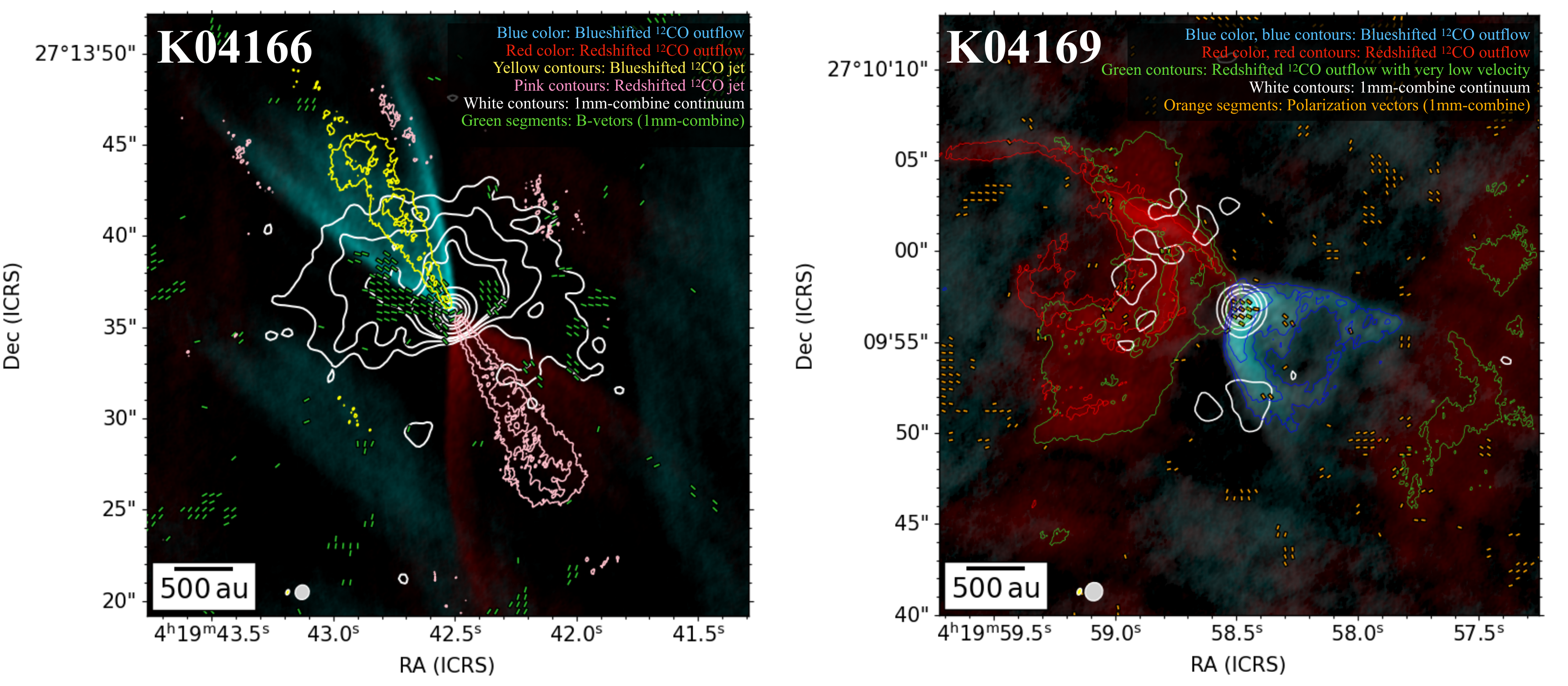}
      \caption{Dust polarization and the molecular outflows and jets obtained from our observation (presented in future literature).
      Left: Blueshifted $\mathrm{^{12}CO}$ outflow indicated by the background cyan color. This represents the moment-0 map obtained from the 1mm-high $\mathrm{^{12}CO}$(2-1) line data (-7 to 5\,$\mathrm{km\,s^{-1}}$). The background red color indicates the redshifted $\mathrm{^{12}CO}$ outflow, corresponding to the same data, but across a different velocity range (7 to 20\,$\mathrm{km\,s^{-1}}$).  
      The yellow contours represent the blueshifted $\mathrm{^{12}CO}$ jet, obtained from the same data (-50 to -35\,$\mathrm{km\,s^{-1}}$) at 30\% and 60\% of the maximum value of the moment-0 map. In contrast, the pink contours depict the redshifted $\mathrm{^{12}CO}$ jet, from the same data with a velocity range of 45 to 58\,$\mathrm{km\,s^{-1}}$, at 30\% and 60\% levels of the maximum value.
      The white contours illustrate the 1mm-combine continuum emission at levels of [3, 5, 8, 10, 30, 100, 200, 400, 700]$\times89\,\mathrm{\mu Jy\,beam}^{-1}$. 
      We adopted the systemic velocity of 6.7\,$\mathrm{km\,s^{-1}}$ \citep{Tafalla_2017A&A...597A.119T}. The green segments indicate the B-vectors from the 1mm-combine map (as shown in panel b of Fig.~\ref{result_K66_polari}).
      Right: Blueshifted $\mathrm{^{12}CO}$ outflow with a moment-0 map from the 1mm-high $\mathrm{^{12}CO}$(2-1) data in a velocity range of -3 to 6\,$\mathrm{km\,s^{-1}}$, again represented by the background . The background red color displays the redshifted $\mathrm{^{12}CO}$ outflow, with data collected in a velocity range of 7 to 16\,$\mathrm{km\,s^{-1}}$. 
      The blue contours highlight the blueshifted $\mathrm{^{12}CO}$ outflow, derived from the same data with a velocity range of -3 to 6\,$\mathrm{km\,s^{-1}}$, with contour levels set at 10\%, 20\%, and 40\% of the maximum value. The red contours depict the redshifted $\mathrm{^{12}CO}$ outflow for the velocity range of 11 to 16\,$\mathrm{km\,s^{-1}}$, with contour levels at 20\% and 50\% of the maximum value. Additionally, the green contours represent the redshifted low-velocity outflow, obtained from data with a velocity range of 7.5 to 10\,$\mathrm{km\,s^{-1}}$, with contour levels of 20\% and 40\% of the maximum value.
      We adopted the systemic velocity of 6.9\,$\mathrm{km\,s^{-1}}$ \citep{Han_2025ApJ...993..120H}.
      The white contours display the 1mm combined continuum emission at levels of [3, 30, 100, 400, 700]$\times116\,\mathrm{\mu Jy\,beam}^{-1}$, while the orange segments correspond to the polarization vectors obtained from the 1mm-combine data, as shown in panel b of Fig.~\ref{result_K69_polari}.
      In both maps, the yellow and gray ellipses at the bottom left present the synthesized beam of the $\mathrm{^{12}CO}$ data ($0.38\arcsec\times0.23\arcsec,$ -23\degr) and the 1mm-combine data, respectively. 
      }
         \label{fig:outflow_polari}
   \end{figure*}

\begin{figure*}
\sidecaption
  \includegraphics[width=12.5cm]{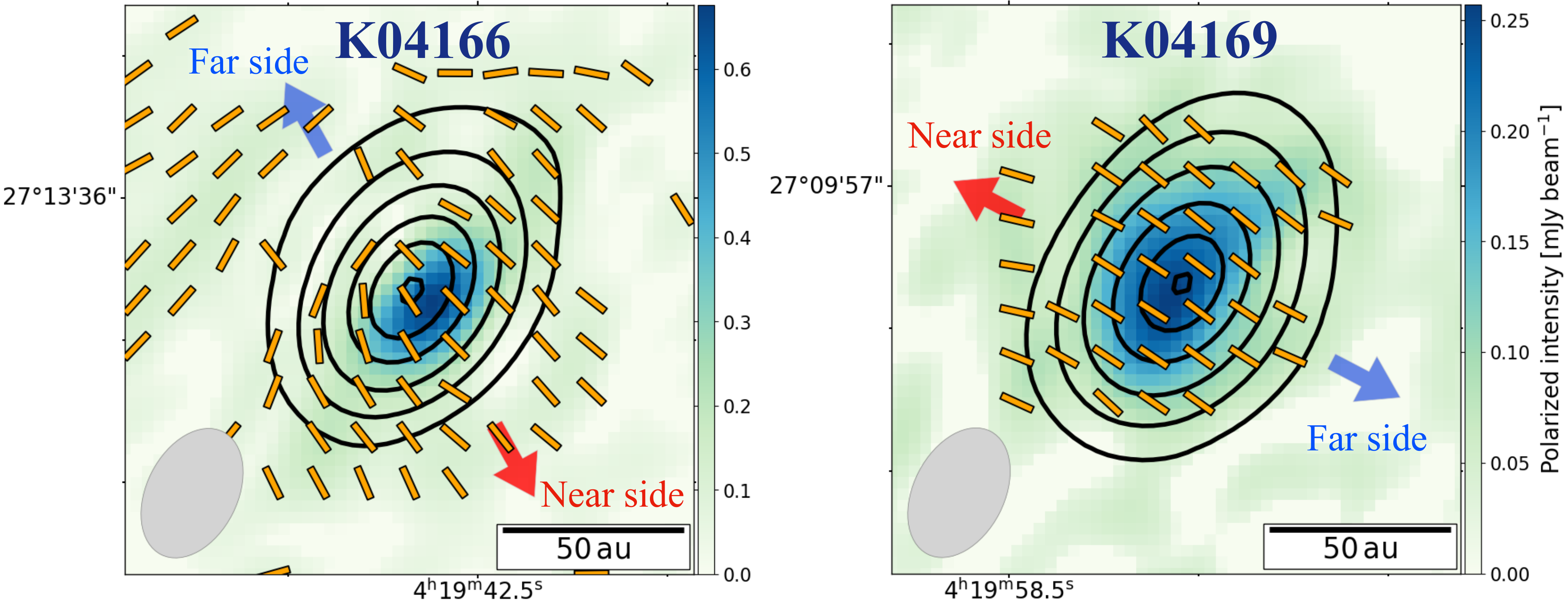}
  \caption{Disk nearside-farside asymmetry between total and polarized intensities obtained from 1mm-high data. Left:\ Polarized intensity map of K04166 in color with the total intensity (the black contours at [50, 200, 500, 1000, 1500, 2000]$\times25\,\mathrm{\mu Jy\,beam}^{-1}$ levels) and polarization vectors (the orange segments). The blue and red arrows and the gray solid ellipse are the same as Fig.~\ref{fig:K66_dis_1}. Right: Polarized intensity map of K04169 in color with the total intensity (the black contours at [50, 200, 500, 1000, 1500, 1850]$\times28\,\mathrm{\mu Jy\,beam}^{-1}$ levels) and polarization vectors (the orange segments). The blue and red arrows and the gray solid ellipse are the same as Fig.~\ref{result_K69_polari}.}
  \label{fig:dis_near_far}
\end{figure*}

K04166 and K04169 are both embedded in the B213 filament of Taurus and it is often assumed they are similarly young protostars formed from the same parent molecular structure, thus sharing similar large-scale environments and initial conditions. At the core scale ($\sim$0.1\,pc), single-dish IRAM\,30m and NIKA observations show that both sources are associated with extended dust emission \citep{Bracco2017A&A...604A..52B}. 
Yet, the maps reveal that K04169 exhibits a more compact dust continuum emission than K04166, despite both sources showing similarly extended emission at the $\sim$0.1\,pc scale in the single-dish filament map (Fig.~\ref{fig:NIKA-ALMA}). Our analysis of the visibility reveals that there is no significant envelope emission associated with K04169 in the ALMA/ACA data; rather, there is a contribution from a diffuse extended component present, which does not contribute to the flux at scales 100-1000 au around the disk. This behavior is more characteristic of cloud emission than envelope emission. This suggests that the surrounding material in K04169 has largely been accreted or dispersed, while K04166 remains embedded within its natal envelope. 
We also evaluated the flux recovered by our ALMA 1mm-combine continuum map, from the single-dish peak flux measured by the IRAM\,30\,m/MAMBO data at 240\,GHz (12\arcsec resolution, \citealt{Motte2001}). We find 90.0\% and 57.2\% of the MAMBO peak flux is recovered at the scales probed by ALMA ($1-7\arcsec$), for K04166 and K04169 respectively.\footnote{\citet{Motte2001} measured the peak flux to be $180\pm8\,\mathrm{mJy\,beam^{-1}}$ for K04166 and $190\pm9\,\mathrm{mJy\,beam^{-1}}$ for K04169 at 240\,GHz in the MAMBO map. We scaled the MAMBO flux to the ALMA frequency (231.8\,GHz) using a spectral index $\alpha = 3$ (as measured at large scales in the NIKA data; see \citealt{Bracco2017A&A...604A..52B}). Note: the ALMA integrated fluxes ($\geq 3\sigma$) come from scales smaller than the 12\arcsec single-dish beam.}
A comparison of Fig.~\ref{fig:NIKA-ALMA} (b) and (d) indicates that the emission associated with K04169 is very compact, while the extended emission recovered by ALMA in the 1mm-combine maps toward K04166 almost fills the 12-arcsec single-dish beam.
This suggests that the missing flux toward K04169 does not originate from its immediate surroundings, but instead from the dust present on a larger scale ($\gtrsim 1000$~au), likely stemming from the B123 filament. Consequently, a substantial portion of the spectral energy distribution (SED) for K04169 at FIR wavelengths may not be linked to this YSO \citep[See Fig.~9 of][]{Ohashi_2023ApJ...951....8O}.
In summary, the dust distribution at the envelope scales indicates that K04166 is a true protostar, while K04169 is significantly more evolved. The lack of an envelope suggests that it is likely in a late Class I  or an early Class II stage.

The protostellar nature of K04166 is also supported by the facts that the envelope is threaded by a well-organized, hourglass-shaped magnetic field, extending down to approximately the 20\,au disk level, and that it shows a notable alignment with the large-scale magnetic field of its surrounding cloud (Fig.~\ref{result_K66_polari}a).
Another protostellar signature is the single collimated outflow and jet toward K04166; while the outflow opening angle at small scales for K04169 is much wider (see Fig.~\ref{fig:outflow_polari}; \citealt{Tafalla_2004A&A...423L..21T}). This is consistent with a later evolutionary phase for K04169, as it has been suggested that the molecular outflows become less collimated when the protostar evolves from the Class\,0 to I stage \citep{Hsieh_2023ApJ...947...25H, Dunham_2024MNRAS.533.3828D}. Finally, while SiO emission is a good tracer of high-velocity shocked gas \citep[e.g.,][]{Mikami_1992ApJ...392L..87M, Sato_2023ApJ...958..102S},  only  K04166 has indicated the presence of an SiO jet so far \citep{Phuong_2025ApJ...992...18P}. All these properties point toward an interpretation of K04166 being significantly younger than K04169.
Finally, at disk scales, both the polarized and total intensities show a pronounced nearside-farside asymmetry along the disk minor axis (Fig.~\ref{fig:dis_near_far}). 
The asymmetry in the total intensity occurs due to geometric effects, as seen in both K04166 and K04169 maps \citep{Takakuwa_2024ApJ...964...24T}.
In contrast, only the K04166 map displays the asymmetry of the polarized intensity along the disk minor axis. This occurs due to the difference in the local inclination angle of the disk surface to the line of sight in optically thick disks, characterizing younger and less settled dusty disks \citep{Yang_2017MNRAS.472..373Y, Bacciotti_2018ApJ...865L..12B}.

In the case of K04169, it might be unexpected to find it associated with a molecular flow if it is a Class II YSO. Still, ALMA data has revealed a gas "streamer" that connects the disk of K04169 to the larger cloud structures \citep{Han_2025ApJ...993..120H}. This could explain the late, yet significant accretion onto the disk and the ensuing presence of the observed active outflow. Such late accretion systems are frequently observed now that low surface brightness gas can be mapped at high angular resolution in local star-forming regions (see, e.g., the T-Tauri disks harboring streamers in \citealt{Huang_DOtau_2022ApJ...930..171H, Garufi_DGtau_2022A&A...658A.104G, Gupta_2023A&A...670L...8G, Huang_DRtau_2023ApJ...943..107H}).
Interestingly, the extent of the disk from the 1mm dust thermal emission is found to be larger for K04169 than for K04166, while the star is more evolved, and both probably formed in similar conditions. We stress that this is not what would be expected by disk evolution models, where large dust grains drift rapidly toward the inner radii, producing smaller dusty disk sizes as the disk evolves \citep{Sanchis_2021A&A...649A..19S, Trapman_2025ApJ...989...10T}.
It has been suggested that a significant fraction of Class I protostars in Taurus may be embedded Class II \citep{Carney2016}. If K04169 is a T-Tauri formed elsewhere, considering typical proper motions of T-Tauri objects (a few km/s), it could have traveled up to 2pc in 1Myrs and thus could have been formed anywhere in the filament, independently of K04166.
The age and spatial extent of B213 \citep[3\,Myrs;][]{Gaia_Krolikowski_2021AJ....162..110K} would thus allow for different evolutionary stages to be found within the local environment.
However, both YSOs are embedded in the B213 filament: they may thus have been formed locally from the same filamentary material. The both large-scale outflows are aligned with the magnetic field lines threading the filament (see \citealt{Palmeirim_2013A&A...550A..38P,Narayanan_2012MNRAS.425.2641N}) and the disks planes show a similar orientation with respect to the surrounding filament (P.A. (125\degr for K04166 and 139\degr for K04169, see Table~\ref{tab:imfit_results}), roughly parallel to the filament elongation and perpendicular to the filament-scale magnetic field. These observed properties further support a scenario of local formation for both YSOs. We note that the antiparallel character of the red and blue lobes results from a difference in the inclination angles of the disks (see the inclination angles derived from aspect ratios in Table~\ref{tab:imfit_results}), generating different projections of the outflowing gas velocities on the line of sight. This could also be due to the initial magnetic field at disk scales being misaligned with the disk rotation axis \citep{Hirano_2019MNRAS.485.4667H}. 
Hence, our data are consistent with a scenario of sequential star formation where B213 had fragmented into cores \citep{Hacar_2013A&A...554A..55H}; however, this process did not occur all at once: K04169 was formed first and is now in a stage of late Class I or early Class II, while K04166 formed second and is now still a Class 0 protostar in its main accretion phase. We note that between the two YSOs, there is a prestellar core, which is possibly the next protostar to be formed in B213, likely within a few $10^4$ years, strengthening the plausibility of a scenario of sequential star formation in B213.

\section{Conclusions}
\label{sec_5}

In this paper, we present high-angular resolution ALMA polarimetric observations of two YSOs in the B213 filament of Taurus, K04166 and K04169, probing disk and envelope scales to study the dust thermal emission, polarization mechanisms, and magnetic field morphologies. These deep observations were obtained in Band 3 (3\,mm) and Band 6 (1.4\,mm), using both ACA 7-m and ALMA 12-m arrays to sample spatial scales from $\sim$25 au to 3000 au. We modeled the Stokes $I$ emission with 2D Gaussian fits and by directly fitting visibility profiles to disentangle disk and envelope contributions. Polarization fractions and orientations were analyzed to distinguish between magnetically aligned grains and self-scattering mechanisms across scales.
While K04166 and K04169 are two known examples for studying how protostars with similar birth environments can grow and evolve, our data shed light on very different small-scale properties for these two neighboring sources. We summarize our main findings below. 
\begin{itemize}
\item{Disk and envelope evolution distinguishes the twin sources: detailed visibility modeling reveals that K04166 retains a massive envelope and a small disk, consistent with a young, actively accreting protostar. K04169 exhibits no significant envelope at scales of 20–3000 au, and its emission is disk-dominated, indicating that it is likely a late Class I or early Class II object. Despite sharing a similar environment, K04166 and K04169  display striking differences, suggesting different episodes of star formation or a different in the delay resulting from the effect of the local magnetic fields on the collapse of dense cores into YSOs (even close by, i.e., within a single filament).}
\item{Multiscale transition in polarization mechanisms: K04166 shows a clear transition from magnetically aligned grain polarization in the envelope (suggesting an hourglass-shaped magnetic field) to self-scattering-induced polarization at disk scales, while K04169 is dominated by self-scattering consistent with large dust grains in its compact disk. We show tentative evidence that some of the 1\,mm polarized emission in K04166, outside the disk radius, might either be due to self-scattering coming from the inner envelope or it traces a toroidal magnetic field surrounding the disk, at scales $\simlt 100$\,au.}
\item{Evidence for early dust grain growth: while we refrained from interpreting these data to propose an associated dust grain size, as the relationship between grain size and self-scattering signature at mm wavelengths is subject to large uncertainties, we stress that large grains are already present in both disks, confirming the scenario of early dust evolution in disks from the Class 0 stage. Moreover, no self-scattering is detected at 3mm from the young K04166 disk, while compact polarization, probably due to self-scattering, is detected at 3mm from K04169, suggesting a more advanced dust evolution in the K04169 disk.}
\end{itemize}

\begin{acknowledgements}
This work is funded by the European Research Council (ERC) under the European Union’s Horizon 2020 research and innovation program (grant agreement No. 101098309 - PEBBLES). AS acknowledges support by the grant CNS2023-145240 (Consolida).
JMG and AS acknowledge support by the
grant PID2023-146675NB-I00
(MCI-AEI-FEDER, UE). This work is also partially supported by the
program Unidad de Excelencia María de Maeztu CEX2020-001058-M. Andrea Bracco acknowledges financial support from the INAF initiative "IAF Astronomy Fellowships in Italy" (grant name MEGASKAT).
\end{acknowledgements}

\bibliographystyle{aa}
\bibliography{aa}

\begin{thebibliography}{95}
\expandafter\ifx\csname natexlab\endcsname\relax\def\natexlab#1{#1}\fi

\bibitem[{{Andersson} {et~al.}(2015){Andersson}, {Lazarian}, \& {Vaillancourt}}]{Andersson_2015ARA&A..53..501A}
{Andersson}, B.~G., {Lazarian}, A., \& {Vaillancourt}, J.~E. 2015, \araa, 53, 501

\bibitem[{{André} {et~al.}(2000){André}, {Ward-Thompson}, \& {Barsony}}]{AWB2000}
{André}, P., {Ward-Thompson}, D., \& {Barsony}, M. 2000, in Protostars and Planets IV, ed. V.~{Mannings}, A.~P. {Boss}, \& S.~S. {Russell}, 59

\bibitem[{{Bacciotti} {et~al.}(2018){Bacciotti}, {Girart}, {Padovani}, {Podio}, {Paladino}, {Testi}, {Bianchi}, {Galli}, {Codella}, {Coffey}, {Favre}, \& {Fedele}}]{Bacciotti_2018ApJ...865L..12B}
{Bacciotti}, F., {Girart}, J.~M., {Padovani}, M., {et~al.} 2018, \apjl, 865, L12

\bibitem[{{Bhandare} {et~al.}(2024){Bhandare}, {Commer{\c{c}}on}, {Laibe}, {Flock}, {Kuiper}, {Henning}, {Mignone}, \& {Marleau}}]{Bhandare2024}
{Bhandare}, A., {Commer{\c{c}}on}, B., {Laibe}, G., {et~al.} 2024, \aap, 687, A158

\bibitem[{{Bontemps} {et~al.}(1996){Bontemps}, {Andre}, {Terebey}, \& {Cabrit}}]{Bontemps_1996A&A...311..858B}
{Bontemps}, S., {Andre}, P., {Terebey}, S., \& {Cabrit}, S. 1996, \aap, 311, 858

\bibitem[{{Bracco} {et~al.}(2017){Bracco}, {Palmeirim}, {Andr{\'e}}, {Adam}, {Ade}, {Bacmann}, {Beelen}, {Beno{\^\i}t}, {Bideaud}, {Billot}, {Bourrion}, {Calvo}, {Catalano}, {Coiffard}, {Comis}, {D'Addabbo}, {D{\'e}sert}, {Didelon}, {Doyle}, {Goupy}, {K{\"o}nyves}, {Kramer}, {Lagache}, {Leclercq}, {Mac{\'\i}as-P{\'e}rez}, {Maury}, {Mauskopf}, {Mayet}, {Monfardini}, {Motte}, {Pajot}, {Pascale}, {Peretto}, {Perotto}, {Pisano}, {Ponthieu}, {Rev{\'e}ret}, {Rigby}, {Ritacco}, {Rodriguez}, {Romero}, {Roy}, {Ruppin}, {Schuster}, {Sievers}, {Triqueneaux}, {Tucker}, \& {Zylka}}]{Bracco2017A&A...604A..52B}
{Bracco}, A., {Palmeirim}, P., {Andr{\'e}}, P., {et~al.} 2017, \aap, 604, A52

\bibitem[{{Brunngr{\"a}ber} \& {Wolf}(2021)}]{Brunngraber2021A&A}
{Brunngr{\"a}ber}, R. \& {Wolf}, S. 2021, \aap, 648, A87

\bibitem[{{Cacciapuoti} {et~al.}(2025){Cacciapuoti}, {Testi}, {Maury}, {Chandler}, {Sakai}, {Ceccarelli}, {Codella}, {De Simone}, {Podio}, {Sabatini}, {Bianchi}, {Macias}, {Miotello}, {Toci}, {Loinard}, {Johnstone}, {Liu}, {Aikawa}, {Shirley}, {Svoboda}, {Sakai}, {Hirota}, {Viti}, {Lefloch}, {Oya}, {Ohashi}, {Feng}, {Fontani}, {Watanabe}, {Lopez-Sepulcre}, {Zhang}, {Vastel}, {Segura-Cox}, {Pineda}, {Isella}, {Klessen}, {Hennebelle}, {Molinari}, \& {Yamamoto}}]{Cacciapuoti2025}
{Cacciapuoti}, L., {Testi}, L., {Maury}, A.~J., {et~al.} 2025, \aap, 700, A188

\bibitem[{Cacciapuoti {et~al.}(2024)Cacciapuoti, Testi, Podio, Codella, Maury, De~Simone, Hennebelle, Lebreuilly, Klessen, \& Molinari}]{Cacciapuoti_2024}
Cacciapuoti, L., Testi, L., Podio, L., {et~al.} 2024, The Astrophysical Journal, 961, 90

\bibitem[{{Carney} {et~al.}(2016){Carney}, {Y{\i}ld{\i}z}, {Mottram}, {van Dishoeck}, {Ramchandani}, \& {J{\o}rgensen}}]{Carney2016}
{Carney}, M.~T., {Y{\i}ld{\i}z}, U.~A., {Mottram}, J.~C., {et~al.} 2016, \aap, 586, A44

\bibitem[{{CASA Team} {et~al.}(2022){CASA Team}, {Bean}, {Bhatnagar}, {Castro}, {Donovan Meyer}, {Emonts}, {Garcia}, {Garwood}, {Golap}, {Gonzalez Villalba}, {Harris}, {Hayashi}, {Hoskins}, {Hsieh}, {Jagannathan}, {Kawasaki}, {Keimpema}, {Kettenis}, {Lopez}, {Marvil}, {Masters}, {McNichols}, {Mehringer}, {Miel}, {Moellenbrock}, {Montesino}, {Nakazato}, {Ott}, {Petry}, {Pokorny}, {Raba}, {Rau}, {Schiebel}, {Schweighart}, {Sekhar}, {Shimada}, {Small}, {Steeb}, {Sugimoto}, {Suoranta}, {Tsutsumi}, {van Bemmel}, {Verkouter}, {Wells}, {Xiong}, {Szomoru}, {Griffith}, {Glendenning}, \& {Kern}}]{CASA_2022PASP..134k4501C}
{CASA Team}, {Bean}, B., {Bhatnagar}, S., {et~al.} 2022, \pasp, 134, 114501

\bibitem[{{Chapman} {et~al.}(2011){Chapman}, {Goldsmith}, {Pineda}, {Clemens}, {Li}, \& {Kr{\v{c}}o}}]{Chapman_2011ApJ...741...21C}
{Chapman}, N.~L., {Goldsmith}, P.~F., {Pineda}, J.~L., {et~al.} 2011, \apj, 741, 21

\bibitem[{{Cutri} {et~al.}(2003){Cutri}, {Skrutskie}, {van Dyk}, {Beichman}, {Carpenter}, {Chester}, {Cambresy}, {Evans}, {Fowler}, {Gizis}, {Howard}, {Huchra}, {Jarrett}, {Kopan}, {Kirkpatrick}, {Light}, {Marsh}, {McCallon}, {Schneider}, {Stiening}, {Sykes}, {Weinberg}, {Wheaton}, {Wheelock}, \& {Zacarias}}]{Cutri_2003yCat.2246....0C}
{Cutri}, R.~M., {Skrutskie}, M.~F., {van Dyk}, S., {et~al.} 2003, {VizieR Online Data Catalog: 2MASS All-Sky Catalog of Point Sources (Cutri+ 2003)}, VizieR On-line Data Catalog: II/246. Originally published in: University of Massachusetts and Infrared Processing and Analysis Center, (IPAC/California Institute of Technology) (2003)

\bibitem[{{Cutri} {et~al.}(2012){Cutri}, {Wright}, {Conrow}, {Bauer}, {Benford}, {Brandenburg}, {Dailey}, {Eisenhardt}, {Evans}, {Fajardo-Acosta}, {Fowler}, {Gelino}, {Grillmair}, {Harbut}, {Hoffman}, {Jarrett}, {Kirkpatrick}, {Leisawitz}, {Liu}, {Mainzer}, {Marsh}, {Masci}, {McCallon}, {Padgett}, {Ressler}, {Royer}, {Skrutskie}, {Stanford}, {Wyatt}, {Tholen}, {Tsai}, {Wachter}, {Wheelock}, {Yan}, {Alles}, {Beck}, {Grav}, {Masiero}, {McCollum}, {McGehee}, {Papin}, \& {Wittman}}]{Cutri_2012wise.rept....1C}
{Cutri}, R.~M., {Wright}, E.~L., {Conrow}, T., {et~al.} 2012, {Explanatory Supplement to the WISE All-Sky Data Release Products}, Explanatory Supplement to the WISE All-Sky Data Release Products

\bibitem[{{Davis} {et~al.}(2010){Davis}, {Chrysostomou}, {Hatchell}, {Wouterloot}, {Buckle}, {Nutter}, {Fich}, {Brunt}, {Butner}, {Cavanagh}, {Curtis}, {Duarte-Cabral}, {di Francesco}, {Etxaluze}, {Friberg}, {Friesen}, {Fuller}, {Graves}, {Greaves}, {Hogerheijde}, {Johnstone}, {Matthews}, {Matthews}, {Rawlings}, {Richer}, {Roberts}, {Sadavoy}, {Simpson}, {Tothill}, {Tsamis}, {Viti}, {Ward-Thompson}, {White}, \& {Yates}}]{Davis_2010MNRAS.405..759D}
{Davis}, C.~J., {Chrysostomou}, A., {Hatchell}, J., {et~al.} 2010, \mnras, 405, 759

\bibitem[{{Dent} {et~al.}(2019){Dent}, {Pinte}, {Cortes}, {M{\'e}nard}, {Hales}, {Fomalont}, \& {de Gregorio-Monsalvo}}]{Dent2019MNRAS.482L..29D}
{Dent}, W.~R.~F., {Pinte}, C., {Cortes}, P.~C., {et~al.} 2019, \mnras, 482, L29

\bibitem[{{Dunham} {et~al.}(2024){Dunham}, {Stephens}, {Myers}, {Bourke}, {Arce}, {Pokhrel}, {Pineda}, \& {Vargas}}]{Dunham_2024MNRAS.533.3828D}
{Dunham}, M.~M., {Stephens}, I.~W., {Myers}, P.~C., {et~al.} 2024, \mnras, 533, 3828

\bibitem[{{Eswaraiah} {et~al.}(2021){Eswaraiah}, {Li}, {Furuya}, {Hasegawa}, {Ward-Thompson}, {Qiu}, {Ohashi}, {Pattle}, {Sadavoy}, {Hull}, {Berry}, {Doi}, {Ching}, {Lai}, {Wang}, {Koch}, {Kwon}, {Kwon}, {Bastien}, {Arzoumanian}, {Coud{\'e}}, {Soam}, {Fanciullo}, {Yen}, {Liu}, {Hoang}, {Ping Chen}, {Shimajiri}, {Liu}, {Chen}, {Li}, {Lyo}, {Hwang}, {Johnstone}, {Rao}, {Bich Ngoc}, {Ngoc Diep}, {Mairs}, {Parsons}, {Tamura}, {Tahani}, {Vivien Chen}, {Nakamura}, {Shinnaga}, {Tang}, {Cho}, {Won Lee}, {Inutsuka}, {Inoue}, {Iwasaki}, {Qian}, {Xie}, {Li}, {Liu}, {Zhang}, {Chen}, {Zhang}, {Zhu}, {Zhou}, {Andr{\'e}}, {Liu}, {Yuan}, {Lu}, {Peretto}, {Bourke}, {Byun}, {Dai}, {Duan}, {Duan}, {Eden}, {Matthews}, {Fiege}, {Fissel}, {Kim}, {Lee}, {Kim}, {Pyo}, {Choi}, {Choi}, {Chrysostomou}, {Jung Chung}, {Ngoc Tram}, {Franzmann}, {Friberg}, {Friesen}, {Fuller}, {Gledhill}, {Graves}, {Greaves}, {Griffin}, {Gu}, {Han}, {Hatchell}, {Hayashi}, {Houde}, {Kawabata}, {Jeong}, {Kang}, {Kang}, {Kang}, {Kataoka}, {Kemper},
  {Rawlings}, {Rawlings}, {Retter}, {Richer}, {Rigby}, {Saito}, {Savini}, {Scaife}, {Seta}, {Kim}, {Hee Kim}, {Kim}, {Kirchschlager}, {Kirk}, {Kobayashi}, {Konyves}, {Kusune}, {Lacaille}, {Law}, {Lee}, {Lee}, {Matsumura}, {Moriarty-Schieven}, {Nagata}, {Nakanishi}, {Onaka}, {Park}, {Tang}, {Tomisaka}, {Tsukamoto}, {Viti}, {Wang}, {Whitworth}, {Yoo}, {Yun}, {Zenko}, {Zhang}, {de Looze}, {Dowell}, {Eyres}, {Falle}, {Robitaille}, \& {van Loo}}]{Chakali_2021ApJ...912L..27E}
{Eswaraiah}, C., {Li}, D., {Furuya}, R.~S., {et~al.} 2021, \apjl, 912, L27

\bibitem[{{Frau} {et~al.}(2011){Frau}, {Galli}, \& {Girart}}]{Frau_2011A&A...535A..44F}
{Frau}, P., {Galli}, D., \& {Girart}, J.~M. 2011, \aap, 535, A44

\bibitem[{{Galametz} {et~al.}(2019){Galametz}, {Maury}, {Valdivia}, {Testi}, {Belloche}, \& {Andr{\'e}}}]{Galametz2019}
{Galametz}, M., {Maury}, A.~J., {Valdivia}, V., {et~al.} 2019, \aap, 632, A5

\bibitem[{{Garufi} {et~al.}(2022){Garufi}, {Podio}, {Codella}, {Segura-Cox}, {Vander Donckt}, {Mercimek}, {Bacciotti}, {Fedele}, {Kasper}, {Pineda}, {Humphreys}, \& {Testi}}]{Garufi_DGtau_2022A&A...658A.104G}
{Garufi}, A., {Podio}, L., {Codella}, C., {et~al.} 2022, \aap, 658, A104

\bibitem[{{Giang} \& {Hoang}(2024)}]{Giang2024}
{Giang}, N.~C. \& {Hoang}, T. 2024, \mnras, 530, 984

\bibitem[{{Giang} {et~al.}(2025){Giang}, {Le Gouellec}, {Hoang}, {Maury}, \& {Hennebelle}}]{Giang2025}
{Giang}, N.~C., {Le Gouellec}, V.~J.~M., {Hoang}, T., {Maury}, A.~J., \& {Hennebelle}, P. 2025, \apj, 980, 105

\bibitem[{{Girart} {et~al.}(2006){Girart}, {Rao}, \& {Marrone}}]{Girart_2006Sci...313..812G}
{Girart}, J.~M., {Rao}, R., \& {Marrone}, D.~P. 2006, Science, 313, 812

\bibitem[{{Goldsmith} {et~al.}(2008){Goldsmith}, {Heyer}, {Narayanan}, {Snell}, {Li}, \& {Brunt}}]{Goldsmith_2008ApJ...680..428G}
{Goldsmith}, P.~F., {Heyer}, M., {Narayanan}, G., {et~al.} 2008, \apj, 680, 428

\bibitem[{{G{\"u}del} {et~al.}(2007){G{\"u}del}, {Briggs}, {Arzner}, {Audard}, {Bouvier}, {Feigelson}, {Franciosini}, {Glauser}, {Grosso}, {Micela}, {Monin}, {Montmerle}, {Padgett}, {Palla}, {Pillitteri}, {Rebull}, {Scelsi}, {Silva}, {Skinner}, {Stelzer}, \& {Telleschi}}]{Gudel_2007A&A...468..353G}
{G{\"u}del}, M., {Briggs}, K.~R., {Arzner}, K., {et~al.} 2007, \aap, 468, 353

\bibitem[{{Gupta} {et~al.}(2023){Gupta}, {Miotello}, {Manara}, {Williams}, {Facchini}, {Beccari}, {Birnstiel}, {Ginski}, {Hacar}, {K{\"u}ffmeier}, {Testi}, {Tychoniec}, \& {Yen}}]{Gupta_2023A&A...670L...8G}
{Gupta}, A., {Miotello}, A., {Manara}, C.~F., {et~al.} 2023, \aap, 670, L8

\bibitem[{{Hacar} {et~al.}(2013){Hacar}, {Tafalla}, {Kauffmann}, \& {Kov{\'a}cs}}]{Hacar_2013A&A...554A..55H}
{Hacar}, A., {Tafalla}, M., {Kauffmann}, J., \& {Kov{\'a}cs}, A. 2013, \aap, 554, A55

\bibitem[{{Han} {et~al.}(2025){Han}, {Kwon}, {Aso}, {Ohashi}, {Tobin}, {J{\o}rgensen}, {Takakuwa}, {Looney}, {Aikawa}, {Flores}, {de Gregorio-Monsalvo}, {Koch}, {Lee}, {Lee}, {Li}, {Lin}, {Sai}, {Thieme}, {Williams}, {Gavino}, {Kido}, {Lai}, {Phuong}, {Santamar{\'\i}a-Miranda}, \& {Yen}}]{Han_2025ApJ...993..120H}
{Han}, I., {Kwon}, W., {Aso}, Y., {et~al.} 2025, \apj, 993, 120

\bibitem[{{Harris} {et~al.}(2018){Harris}, {Cox}, {Looney}, {Li}, {Yang}, {Fern{\'a}ndez-L{\'o}pez}, {Kwon}, {Sadavoy}, {Segura-Cox}, {Stephens}, \& {Tobin}}]{Harris2018}
{Harris}, R.~J., {Cox}, E.~G., {Looney}, L.~W., {et~al.} 2018, \apj, 861, 91

\bibitem[{{Hennebelle} \& {Fromang}(2008)}]{Hennebelle_2008A&A...477....9H}
{Hennebelle}, P. \& {Fromang}, S. 2008, \aap, 477, 9

\bibitem[{{Hirano} \& {Machida}(2019)}]{Hirano_2019MNRAS.485.4667H}
{Hirano}, S. \& {Machida}, M.~N. 2019, \mnras, 485, 4667

\bibitem[{{Hoang} {et~al.}(2022){Hoang}, {Tram}, {Minh Phan}, {Giang}, {Phuong}, \& {Dieu}}]{Hoang2022}
{Hoang}, T., {Tram}, L.~N., {Minh Phan}, V.~H., {et~al.} 2022, \aj, 164, 248

\bibitem[{{Hsieh} {et~al.}(2023){Hsieh}, {Arce}, {Li}, {Dunham}, {Offner}, {Stephens}, {Stutz}, {Megeath}, {Kong}, {Plunkett}, {Tobin}, {Zhang}, {Mardones}, {Pineda}, {Stanke}, \& {Carpenter}}]{Hsieh_2023ApJ...947...25H}
{Hsieh}, C.-H., {Arce}, H.~G., {Li}, Z.-Y., {et~al.} 2023, \apj, 947, 25

\bibitem[{{Huang} {et~al.}(2024){Huang}, {Girart}, {Stephens}, {Fern{\'a}ndez L{\'o}pez}, {Arce}, {Carpenter}, {Cortes}, {Cox}, {Friesen}, {Le Gouellec}, {Hull}, {Karnath}, {Kwon}, {Li}, {Looney}, {Megeath}, {Myers}, {Murillo}, {Pineda}, {Sadavoy}, {S{\'a}nchez-Monge}, {Sanhueza}, {Tobin}, {Zhang}, {Jackson}, \& {Segura-Cox}}]{Huang_2024ApJ...963L..31H}
{Huang}, B., {Girart}, J.~M., {Stephens}, I.~W., {et~al.} 2024, \apjl, 963, L31

\bibitem[{{Huang} {et~al.}(2023){Huang}, {Bergin}, {Bae}, {Benisty}, \& {Andrews}}]{Huang_DRtau_2023ApJ...943..107H}
{Huang}, J., {Bergin}, E.~A., {Bae}, J., {Benisty}, M., \& {Andrews}, S.~M. 2023, \apj, 943, 107

\bibitem[{{Huang} {et~al.}(2022){Huang}, {Ginski}, {Benisty}, {Ren}, {Bohn}, {Choquet}, {{\"O}berg}, {Ribas}, {Bae}, {Bergin}, {Birnstiel}, {Boehler}, {Facchini}, {Harsono}, {Hogerheijde}, {Long}, {Manara}, {M{\'e}nard}, {Pinilla}, {Pinte}, {Rab}, {Williams}, \& {Zurlo}}]{Huang_DOtau_2022ApJ...930..171H}
{Huang}, J., {Ginski}, C., {Benisty}, M., {et~al.} 2022, \apj, 930, 171

\bibitem[{{Hull} {et~al.}(2020){Hull}, {Cortes}, {Gouellec}, {Girart}, {Nagai}, {Nakanishi}, {Kameno}, {Fomalont}, {Brogan}, {Moellenbrock}, {Paladino}, \& {Villard}}]{Hull_2020PASP..132i4501H}
{Hull}, C. L.~H., {Cortes}, P.~C., {Gouellec}, V. J.~M.~L., {et~al.} 2020, \pasp, 132, 094501

\bibitem[{{Hull} \& {Plambeck}(2015)}]{Hull2015JAI.....450005H}
{Hull}, C. L.~H. \& {Plambeck}, R.~L. 2015, Journal of Astronomical Instrumentation, 4, 1550005

\bibitem[{{J{\o}rgensen} {et~al.}(2007){J{\o}rgensen}, {Bourke}, {Myers}, {Di Francesco}, {van Dishoeck}, {Lee}, {Ohashi}, {Sch{\"o}ier}, {Takakuwa}, {Wilner}, \& {Zhang}}]{Jorgensen2007}
{J{\o}rgensen}, J.~K., {Bourke}, T.~L., {Myers}, P.~C., {et~al.} 2007, \apj, 659, 479

\bibitem[{{Kataoka} {et~al.}(2015){Kataoka}, {Muto}, {Momose}, {Tsukagoshi}, {Fukagawa}, {Shibai}, {Hanawa}, {Murakawa}, \& {Dullemond}}]{Kataoka_2015ApJ...809...78K}
{Kataoka}, A., {Muto}, T., {Momose}, M., {et~al.} 2015, \apj, 809, 78

\bibitem[{{Kataoka} {et~al.}(2017){Kataoka}, {Tsukagoshi}, {Pohl}, {Muto}, {Nagai}, {Stephens}, {Tomisaka}, \& {Momose}}]{Kataoka_2017ApJ...844L...5K}
{Kataoka}, A., {Tsukagoshi}, T., {Pohl}, A., {et~al.} 2017, \apjl, 844, L5

\bibitem[{{Kirchschlager} \& {Bertrang}(2020)}]{Kirchschlager_2020A&A...638A.116K}
{Kirchschlager}, F. \& {Bertrang}, G. H.~M. 2020, \aap, 638, A116

\bibitem[{{Krolikowski} {et~al.}(2021){Krolikowski}, {K1raus}, \& {Rizzuto}}]{Gaia_Krolikowski_2021AJ....162..110K}
{Krolikowski}, D.~M., {K1raus}, A.~L., \& {Rizzuto}, A.~C. 2021, \aj, 162, 110

\bibitem[{{Kwon} {et~al.}(2019){Kwon}, {Stephens}, {Tobin}, {Looney}, {Li}, {van der Tak}, \& {Crutcher}}]{Kwon_2019ApJ...879...25K}
{Kwon}, W., {Stephens}, I.~W., {Tobin}, J.~J., {et~al.} 2019, \apj, 879, 25

\bibitem[{{Lazarian}(2007)}]{Lazarian_2007JQSRT.106..225L}
{Lazarian}, A. 2007, \jqsrt, 106, 225

\bibitem[{{Le Gouellec} {et~al.}(2019){Le Gouellec}, {Hull}, {Maury}, {Girart}, {Tychoniec}, {Kristensen}, {Li}, {Louvet}, {Cortes}, \& {Rao}}]{LeGouellec_2019ApJ...885..106L}
{Le Gouellec}, V. J.~M., {Hull}, C. L.~H., {Maury}, A.~J., {et~al.} 2019, \apj, 885, 106

\bibitem[{{Le Gouellec} {et~al.}(2023){Le Gouellec}, {Maury}, {Hull}, {Verliat}, {Hennebelle}, \& {Valdivia}}]{LeGouellec_2023A&A...675A.133L}
{Le Gouellec}, V.~J.~M., {Maury}, A.~J., {Hull}, C.~L.~H., {et~al.} 2023, \aap, 675, A133

\bibitem[{{Lebreuilly} {et~al.}(2020){Lebreuilly}, {Commer{\c{c}}on}, \& {Laibe}}]{Lebreuilly_2020A&A...641A.112L}
{Lebreuilly}, U., {Commer{\c{c}}on}, B., \& {Laibe}, G. 2020, \aap, 641, A112

\bibitem[{{Lee} {et~al.}(2021){Lee}, {Li}, {Yang}, {Daniel Lin}, {Ching}, \& {Lai}}]{CFLee2021}
{Lee}, C.-F., {Li}, Z.-Y., {Yang}, H., {et~al.} 2021, \apj, 910, 75

\bibitem[{{Liu} {et~al.}(2024){Liu}, {Takahashi}, {Machida}, {Tomisaka}, {Girart}, {Ho}, {Nakanishi}, \& {Sato}}]{Liu_2024ApJ...963..104L}
{Liu}, Y., {Takahashi}, S., {Machida}, M., {et~al.} 2024, \apj, 963, 104

\bibitem[{{Looney} {et~al.}(2000){Looney}, {Mundy}, \& {Welch}}]{Looney2000}
{Looney}, L.~W., {Mundy}, L.~G., \& {Welch}, W.~J. 2000, \apj, 529, 477

\bibitem[{{Machida} {et~al.}(2008){Machida}, {Inutsuka}, \& {Matsumoto}}]{Machida_2008ApJ...676.1088M}
{Machida}, M.~N., {Inutsuka}, S.-i., \& {Matsumoto}, T. 2008, \apj, 676, 1088

\bibitem[{{Marsh} {et~al.}(2014){Marsh}, {Griffin}, {Palmeirim}, {Andr{\'e}}, {Kirk}, {Stamatellos}, {Ward-Thompson}, {Roy}, {Bontemps}, {di Francesco}, {Elia}, {Hill}, {K{\"o}nyves}, {Motte}, {Nguyen-Luong}, {Peretto}, {Pezzuto}, {Rivera-Ingraham}, {Schneider}, {Spinoglio}, \& {White}}]{Marsh_2014MNRAS.439.3683M}
{Marsh}, K.~A., {Griffin}, M.~J., {Palmeirim}, P., {et~al.} 2014, \mnras, 439, 3683

\bibitem[{{Marsh} {et~al.}(2016){Marsh}, {Kirk}, {Andr{\'e}}, {Griffin}, {K{\"o}nyves}, {Palmeirim}, {Men'shchikov}, {Ward-Thompson}, {Benedettini}, {Bresnahan}, {di Francesco}, {Elia}, {Motte}, {Peretto}, {Pezzuto}, {Roy}, {Sadavoy}, {Schneider}, {Spinoglio}, \& {White}}]{Marsh_2016MNRAS.459..342M}
{Marsh}, K.~A., {Kirk}, J.~M., {Andr{\'e}}, P., {et~al.} 2016, \mnras, 459, 342

\bibitem[{{Maury} {et~al.}(2022){Maury}, {Hennebelle}, \& {Girart}}]{Maury_2022FrASS...9.9223M}
{Maury}, A., {Hennebelle}, P., \& {Girart}, J.~M. 2022, Frontiers in Astronomy and Space Sciences, 9, 949223

\bibitem[{{Maury} {et~al.}(2011){Maury}, {Andr{\'e}}, {Men'shchikov}, {K{\"o}nyves}, \& {Bontemps}}]{Maury_2011A&A...535A..77M}
{Maury}, A.~J., {Andr{\'e}}, P., {Men'shchikov}, A., {K{\"o}nyves}, V., \& {Bontemps}, S. 2011, \aap, 535, A77

\bibitem[{{Maury} {et~al.}(2019){Maury}, {Andr{\'e}}, {Testi}, {Maret}, {Belloche}, {Hennebelle}, {Cabrit}, {Codella}, {Gueth}, {Podio}, {Anderl}, {Bacmann}, {Bontemps}, {Gaudel}, {Ladjelate}, {Lef{\`e}vre}, {Tabone}, \& {Lefloch}}]{Maury_2019A&A...621A..76M}
{Maury}, A.~J., {Andr{\'e}}, P., {Testi}, L., {et~al.} 2019, \aap, 621, A76

\bibitem[{{Maury} {et~al.}(2018){Maury}, {Girart}, {Zhang}, {Hennebelle}, {Keto}, {Rao}, {Lai}, {Ohashi}, \& {Galametz}}]{Maury2018}
{Maury}, A.~J., {Girart}, J.~M., {Zhang}, Q., {et~al.} 2018, \mnras, 477, 2760

\bibitem[{{Mikami} {et~al.}(1992){Mikami}, {Umemoto}, {Yamamoto}, \& {Saito}}]{Mikami_1992ApJ...392L..87M}
{Mikami}, H., {Umemoto}, T., {Yamamoto}, S., \& {Saito}, S. 1992, \apjl, 392, L87

\bibitem[{{Mizuno} {et~al.}(1994){Mizuno}, {Onishi}, {Hayashi}, {Ohashi}, {Sunada}, {Hasegawa}, \& {Fukui}}]{Mizuno_1994Natur.368..719M}
{Mizuno}, A., {Onishi}, T., {Hayashi}, M., {et~al.} 1994, \nat, 368, 719

\bibitem[{{Motte} \& {Andr{\'e}}(2001)}]{Motte2001}
{Motte}, F. \& {Andr{\'e}}, P. 2001, \aap, 365, 440

\bibitem[{{Naghizadeh-Khouei} \& {Clarke}(1993)}]{Naghizadeh-Khouei_1993A&A...274..968N}
{Naghizadeh-Khouei}, J. \& {Clarke}, D. 1993, \aap, 274, 968

\bibitem[{{Narayanan} {et~al.}(2012){Narayanan}, {Snell}, \& {Bemis}}]{Narayanan_2012MNRAS.425.2641N}
{Narayanan}, G., {Snell}, R., \& {Bemis}, A. 2012, \mnras, 425, 2641

\bibitem[{{Ohashi} {et~al.}(2023){Ohashi}, {Tobin}, {J{\o}rgensen}, {Takakuwa}, {Sheehan}, {Aikawa}, {Li}, {Looney}, {Williams}, {Aso}, {Sharma}, {Sai}, {Yamato}, {Lee}, {Tomida}, {Yen}, {Encalada}, {Flores}, {Gavino}, {Kido}, {Han}, {Lin}, {Narayanan}, {Phuong}, {Santamar{\'\i}a-Miranda}, {Thieme}, {van't Hoff}, {de Gregorio-Monsalvo}, {Koch}, {Kwon}, {Lai}, {Lee}, {Plunkett}, {Saigo}, {Hirano}, {Lam}, \& {Mori}}]{Ohashi_2023ApJ...951....8O}
{Ohashi}, N., {Tobin}, J.~J., {J{\o}rgensen}, J.~K., {et~al.} 2023, \apj, 951, 8

\bibitem[{{Palmeirim} {et~al.}(2013){Palmeirim}, {Andr{\'e}}, {Kirk}, {Ward-Thompson}, {Arzoumanian}, {K{\"o}nyves}, {Didelon}, {Schneider}, {Benedettini}, {Bontemps}, {Di Francesco}, {Elia}, {Griffin}, {Hennemann}, {Hill}, {Martin}, {Men'shchikov}, {Molinari}, {Motte}, {Nguyen Luong}, {Nutter}, {Peretto}, {Pezzuto}, {Roy}, {Rygl}, {Spinoglio}, \& {White}}]{Palmeirim_2013A&A...550A..38P}
{Palmeirim}, P., {Andr{\'e}}, P., {Kirk}, J., {et~al.} 2013, \aap, 550, A38

\bibitem[{{Pearson} \& {Readhead}(1984)}]{Pearson1984}
{Pearson}, T.~J. \& {Readhead}, A.~C.~S. 1984, \araa, 22, 97

\bibitem[{{Phuong} {et~al.}(2025){Phuong}, {Lee}, {Tobin}, {Ohashi}, {J{\o}rgensen}, {Takakuwa}, {Aikawa}, {Aso}, {Li}, {Koch}, {Williams}, {Gavino}, {Lin}, {Tomida}, {Kwon}, {Looney}, {Han}, {Santamar{\'\i}a-Miranda}, {Lai}, {Yen}, {Thieme}, {Sai}, \& {Flores}}]{Phuong_2025ApJ...992...18P}
{Phuong}, N.~T., {Lee}, C.~W., {Tobin}, J.~J., {et~al.} 2025, \apj, 992, 18

\bibitem[{Rebull {et~al.}(2010)Rebull, Padgett, McCabe, Hillenbrand, Stapelfeldt, Noriega-Crespo, Carey, Brooke, Huard, Terebey, Audard, Monin, Fukagawa, Güdel, Knapp, Menard, Allen, Angione, Baldovin-Saavedra, Bouvier, Briggs, Dougados, Evans, Flagey, Guieu, Grosso, Glauser, Harvey, Hines, Latter, Skinner, Strom, Tromp, \& Wolf}]{Rebull_2010}
Rebull, L.~M., Padgett, D.~L., McCabe, C.-E., {et~al.} 2010, The Astrophysical Journal Supplement Series, 186, 259

\bibitem[{{Sadavoy} {et~al.}(2019){Sadavoy}, {Stephens}, {Myers}, {Looney}, {Tobin}, {Kwon}, {Commer{\c{c}}on}, {Segura-Cox}, {Henning}, \& {Hennebelle}}]{Sadavoy2019}
{Sadavoy}, S.~I., {Stephens}, I.~W., {Myers}, P.~C., {et~al.} 2019, \apjs, 245, 2

\bibitem[{{Sanchis} {et~al.}(2021){Sanchis}, {Testi}, {Natta}, {Facchini}, {Manara}, {Miotello}, {Ercolano}, {Henning}, {Preibisch}, {Carpenter}, {de Gregorio-Monsalvo}, {Jayawardhana}, {Lopez}, {Mu{\v{z}}i{\'c}}, {Pascucci}, {Santamar{\'\i}a-Miranda}, {van Terwisga}, \& {Williams}}]{Sanchis_2021A&A...649A..19S}
{Sanchis}, E., {Testi}, L., {Natta}, A., {et~al.} 2021, \aap, 649, A19

\bibitem[{{Sato} {et~al.}(2023){Sato}, {Tokuda}, {Machida}, {Tachihara}, {Harada}, {Yamasaki}, {Hirano}, {Onishi}, \& {Matsushita}}]{Sato_2023ApJ...958..102S}
{Sato}, A., {Tokuda}, K., {Machida}, M.~N., {et~al.} 2023, \apj, 958, 102

\bibitem[{{Schmalzl} {et~al.}(2010){Schmalzl}, {Kainulainen}, {Quanz}, {Alves}, {Goodman}, {Henning}, {Launhardt}, {Pineda}, \& {Rom{\'a}n-Z{\'u}{\~n}iga}}]{Schmalzl2010}
{Schmalzl}, M., {Kainulainen}, J., {Quanz}, S.~P., {et~al.} 2010, \apj, 725, 1327

\bibitem[{{Serkowski}(1958)}]{Serkowski_1958AcA.....8..135S}
{Serkowski}, K. 1958, \actaa, 8, 135

\bibitem[{{Sheehan} {et~al.}(2022){Sheehan}, {Tobin}, {Looney}, \& {Megeath}}]{Sheehan2022}
{Sheehan}, P.~D., {Tobin}, J.~J., {Looney}, L.~W., \& {Megeath}, S.~T. 2022, \apj, 929, 76

\bibitem[{{Shu} {et~al.}(1987){Shu}, {Adams}, \& {Lizano}}]{Shu1987}
{Shu}, F.~H., {Adams}, F.~C., \& {Lizano}, S. 1987, \araa, 25, 23

\bibitem[{{Soler}(2019)}]{Soler2019}
{Soler}, J.~D. 2019, \aap, 629, A96

\bibitem[{{Soler} {et~al.}(2023){Soler}, {Zucker}, {Peek}, {Heyer}, {Goldsmith}, {Glover}, {Molinari}, {Klessen}, {Hennebelle}, {Testi}, {Colman}, {Benedettini}, {Elia}, {Mininni}, {Pezzuto}, {Schisano}, \& {Traficante}}]{Soler2023}
{Soler}, J.~D., {Zucker}, C., {Peek}, J.~E.~G., {et~al.} 2023, \aap, 675, A206

\bibitem[{{Stephens} {et~al.}(2017){Stephens}, {Yang}, {Li}, {Looney}, {Kataoka}, {Kwon}, {Fern{\'a}ndez-L{\'o}pez}, {Hull}, {Hughes}, {Segura-Cox}, {Mundy}, {Crutcher}, \& {Rao}}]{Stephens_2017ApJ...851...55S}
{Stephens}, I.~W., {Yang}, H., {Li}, Z.-Y., {et~al.} 2017, \apj, 851, 55

\bibitem[{{Tafalla} {et~al.}(2004){Tafalla}, {Santiago}, {Johnstone}, \& {Bachiller}}]{Tafalla_2004A&A...423L..21T}
{Tafalla}, M., {Santiago}, J., {Johnstone}, D., \& {Bachiller}, R. 2004, \aap, 423, L21

\bibitem[{{Tafalla} {et~al.}(2017){Tafalla}, {Su}, {Shang}, {Johnstone}, {Zhang}, {Santiago-Garc{\'\i}a}, {Lee}, {Hirano}, \& {Wang}}]{Tafalla_2017A&A...597A.119T}
{Tafalla}, M., {Su}, Y.~N., {Shang}, H., {et~al.} 2017, \aap, 597, A119

\bibitem[{{Takahashi} {et~al.}(2019){Takahashi}, {Machida}, {Tomisaka}, {Ho}, {Fomalont}, {Nakanishi}, \& {Girart}}]{Takahashi_2019ApJ...872...70T}
{Takahashi}, S., {Machida}, M.~N., {Tomisaka}, K., {et~al.} 2019, \apj, 872, 70

\bibitem[{{Takakuwa} {et~al.}(2024){Takakuwa}, {Saigo}, {Kido}, {Ohashi}, {Tobin}, {J{\o}rgensen}, {Aikawa}, {Aso}, {Gavino}, {Han}, {Koch}, {Kwon}, {Lee}, {Lee}, {Li}, {Lin}, {Looney}, {Mori}, {Sai}, {Sharma}, {Sheehan}, {Tomida}, {Williams}, {Yamato}, \& {Yen}}]{Takakuwa_2024ApJ...964...24T}
{Takakuwa}, S., {Saigo}, K., {Kido}, M., {et~al.} 2024, \apj, 964, 24

\bibitem[{{Takakuwa} {et~al.}(2018){Takakuwa}, {Tsukamoto}, {Saigo}, \& {Saito}}]{Takakuwa_2018ApJ...865...51T}
{Takakuwa}, S., {Tsukamoto}, Y., {Saigo}, K., \& {Saito}, M. 2018, \apj, 865, 51

\bibitem[{{Tatematsu} {et~al.}(2004){Tatematsu}, {Umemoto}, {Kandori}, \& {Sekimoto}}]{Tatematsu_2004ApJ...606..333T}
{Tatematsu}, K., {Umemoto}, T., {Kandori}, R., \& {Sekimoto}, Y. 2004, \apj, 606, 333

\bibitem[{{The ALMA Proposer's}(Guide)}]{position_accuracy}
{The ALMA Proposer's}. Guide, Section A.9.5 and ALMA Technical Handbook, Chapter 10.5.2 Astrometric Observations., \url{https://almascience.eso.org/documents-and-tools/documents-and-tools}, accessed: 2025-07-15

\bibitem[{{Tokuda} {et~al.}(2020){Tokuda}, {Fujishiro}, {Tachihara}, {Takashima}, {Fukui}, {Zahorecz}, {Saigo}, {Matsumoto}, {Tomida}, {Machida}, {Inutsuka}, {Andr{\'e}}, {Kawamura}, \& {Onishi}}]{Tokuda_2020ApJ...899...10T}
{Tokuda}, K., {Fujishiro}, K., {Tachihara}, K., {et~al.} 2020, \apj, 899, 10

\bibitem[{{Trapman} {et~al.}(2025){Trapman}, {Vioque}, {Kurtovic}, {Zhang}, {Rosotti}, {Pinilla}, {Carpenter}, {Cieza}, {Pascucci}, {Anania}, {Agurto-Gangas}, {Deng}, {Miley}, {P{\'e}rez}, {Sierra}, {Tabone}, {Ruiz-Rodriguez}, {Gonz{\'a}lez-Ruilova}, \& {TorresVillanueva}}]{Trapman_2025ApJ...989...10T}
{Trapman}, L., {Vioque}, M., {Kurtovic}, N.~T., {et~al.} 2025, \apj, 989, 10

\bibitem[{Tsukamoto {et~al.}(2021)Tsukamoto, Machida, \& Inutsuka}]{Tsukamoto_2021}
Tsukamoto, Y., Machida, M.~N., \& Inutsuka, S.-i. 2021, The Astrophysical Journal Letters, 920, L35

\bibitem[{{Tsukamoto} {et~al.}(2023){Tsukamoto}, {Maury}, {Commercon}, {Alves}, {Cox}, {Sakai}, {Ray}, {Zhao}, \& {Machida}}]{Tsukamoto_2023ASPC..534..317T}
{Tsukamoto}, Y., {Maury}, A., {Commercon}, B., {et~al.} 2023, in Astronomical Society of the Pacific Conference Series, Vol. 534, Protostars and Planets VII, ed. S.~{Inutsuka}, Y.~{Aikawa}, T.~{Muto}, K.~{Tomida}, \& M.~{Tamura}, 317

\bibitem[{{Tung} {et~al.}(2024){Tung}, {Testi}, {Lebreuilly}, {Hennebelle}, {Maury}, {Klessen}, {Cacciapuoti}, {Gonz{\'a}lez}, {Rosotti}, \& {Molinari}}]{Tung2024}
{Tung}, N.-D., {Testi}, L., {Lebreuilly}, U., {et~al.} 2024, \aap, 684, A36

\bibitem[{{Vaillancourt}(2006)}]{Vaillancourt2006PASP..118.1340V}
{Vaillancourt}, J.~E. 2006, \pasp, 118, 1340

\bibitem[{Wong {et~al.}(2016)Wong, Hirashita, \& Li}]{Wong_2016_10.1093/pasj/psw066}
Wong, Y. H.~V., Hirashita, H., \& Li, Z.-Y. 2016, Publications of the Astronomical Society of Japan, 68, 67

\bibitem[{{Wurster} \& {Li}(2018)}]{Wurster2018}
{Wurster}, J. \& {Li}, Z.-Y. 2018, Frontiers in Astronomy and Space Sciences, 5, 39

\bibitem[{{Yang} {et~al.}(2017){Yang}, {Li}, {Looney}, {Girart}, \& {Stephens}}]{Yang_2017MNRAS.472..373Y}
{Yang}, H., {Li}, Z.-Y., {Looney}, L.~W., {Girart}, J.~M., \& {Stephens}, I.~W. 2017, \mnras, 472, 373

\end{thebibliography}

\begin{appendix}
\onecolumn
\section{Observations and map-making}
The details of the observations carried out are presented in Table\,\ref{obs_para}. The imaging parameters used to produce the maps presented in this analysis are summarized in Table\,\ref{tab:para-continuum}.

\begin{table}
    \centering
    \caption{Observation parameters for the analyzed data.}
    \label{obs_para}
    \small
    \begin{tabular}{l|c|c|c}
        \hline\hline
        Project code & 2019.1.01209.S & 2021.1.01420.S & 2021.1.01420.S \\
        \hline
        Band & 3 & 6 & 6 \\
        Array & ALMA 12-m array & ACA 7-m array & ALMA 12-m array \\
        Configuration & C43-3 & - & C43-6 \\
        Representative frequency (GHz)& 93.17 & 219.55 & 219.55  \\
        Number of executions & 3 & 6 & 2 \\
        Observation date (YYYY-MM-DD) & 2019-11-02 & 2022-05-18 -- 2022-07-03 & 2022-07-13 \\
        Number of antennas & 41 -- 42 & 9 -- 10 & 37 -- 41 \\
        Phase center for K04166 (ICRS) & $04^{\mathrm{h}}19^{\mathrm{m}}42.50^{\mathrm{s}}$, $+27^\circ13'13.38''$ & $04^{\mathrm{h}}19^{\mathrm{m}}42.50^{\mathrm{s}}$, $+27^\circ13'36.10''$ & $04^{\mathrm{h}}19^{\mathrm{m}}42.50^{\mathrm{s}}$, $+27^\circ13'36.10''$ \\
        Phase center for K04169 (ICRS) & $04^{\mathrm{h}}19^{\mathrm{m}}58.46^{\mathrm{s}}$, $+27^\circ09'56.90''$ & $04^{\mathrm{h}}19^{\mathrm{m}}58.46^{\mathrm{s}}$, $+27^\circ09'56.90''$ & $04^{\mathrm{h}}19^{\mathrm{m}}58.46^{\mathrm{s}}$, $+27^\circ09'56.90''$ \\
        HPBW primary beam (arcsec) & 49.7 & 40.8 & 21.6 \\
        Mean PWV (mm) & 3.17 -- 3.43 & 0.53 -- 1.56 & 0.51, 0.61 \\
        Baseline ($k\lambda$) & 4 - 151 & 5 - 30  & 10 - 1744 \\
        Maximum recoverable size (arcsec) & 17.1 & 21.3 -- 30.5 & 2.7 \\
        \hline
    \end{tabular}
    \tablefoot{
        HPBW (full width at half power) of the primary beam is defined as $\sim1.13 \lambda / D$ (ALMA technical handbook).
    }
\end{table}

\begin{table}
    \centering
    \caption{Continuum imaging parameters.}
    \label{tab:para-continuum}
    \small
    \begin{tabular}{lcccccccc}
        \hline\hline
        Source & Project & Dataset & Array & Rest frequency & Robust & Synthesized beam, P.A. & Stokes & Noise level \\ 
        & & && (GHz) & & (arcsec $\times$ arcsec, deg.) & & ($\mathrm{{\mu Jy\,beam}^{-1}}$) \\
        \hline
        K04166 & \#2019.1.01209.S & 3mm & 12-m & 94.5 & 0.5 & $1.96 \times 1.52$, $-$23 & $I$ & 17 \\
        &&&&&&& $Q$ & 16 \\ 
        &&&&&&& $U$ & 16 \\
        \cline{2-9}
        & \#2021.1.01420.S & 1mm-combine & 12-m + 7-m & 230.8 & 0.5 & $1.01 \times 0.95$, $-$27 & $I$ & 89 \\
        &&&&&&& $U$ & 69 \\
        &&&&&&& $Q$ & 44 \\ 
        \cline{3-9}
        && 1mm-high & 12-m & 218.4 & 0.5 & $0.24 \times 0.16$, $-$27 & $I$ & 25 \\
        &&&&&&& $Q$ & 21 \\
        &&&&&&& $U$ & 21 \\
        && 1mm-high & 12-m & 218.4 & $+$2.0 & $0.38 \times 0.23$, $-$24 & $I$ & 33 \\
        &&&&&&& $Q$ & 18 \\
        &&&&&&& $U$ & 18 \\
        \hline
        K04169 & \#2019.1.01209.S & 3mm & 12-m & 94.5 & 0.5 & $1.97 \times 1.51$, $-$26 & $I$ & 18 \\
        &&&&&&& $Q$ & 16 \\ 
        &&&&&&& $U$ & 16 \\ 
        \cline{2-9}
        & \#2021.1.01420.S & 1mm-combine & 12-m + 7-m & 230.8 & 0.5 & $1.02 \times 0.95$, $-$27 &            $I$ & 116 \\
        &&&&&&& $Q$ & 71 \\ 
        &&&&&&& $U$ & 45 \\ 
        \cline{3-9}
        && 1mm-high & 12-m & 218.4 & 0.5 & $0.24 \times 0.16$, $-$27 & $I$ & 28 \\
        &&&&&&& $Q$ & 21 \\ 
        &&&&&&& $U$ & 21 \\
        \hline
    \end{tabular}
    \tablefoot{
     The dataset of 1mm-combine was obtained by combining the ACA 7-m array and ALMA 12-m array data in Band\,6. For the 1mm-high data of K04166, we used the maps with the natural weighting in the result section to clearly distinguish the characteristics of the polarized emission in the inner and outer regions, such as the position angle of the polarization vectors, by imaging the extended emission. In contrast, we used the maps with Briggs weighting in the discussion section to investigate the transition region (borderline) between the inner and outer regions, which corresponds to the disk size. In addition, the visibility fitting requires us to average the channels with similar frequencies for the 1mm-combine data. We, thus, extracted two spectral windows that have similar rest frequencies out of the three spectral windows to make the required visibility. That is why their rest frequency is different from that in the 1mm-high data.
    }
\end{table}

\section{imfit to the polarized intensity maps (K04169)}
To compare the morphologies in the polarized intensity with those in total intensity, we implemented the image-based fitting to the polarization maps of K04169. Table\,\ref{tab:imfit_pol_app} presents the results obtained from 2D Gaussian fitting on the polarization maps for K04169 using \texttt{imfit}.

\begin{table*}
\centering
\small
\caption{2D Gaussian fitting on the polarized intensity maps for K04169.}
\label{tab:imfit_pol_app}
\begin{tabular}{llllllllll}
\hline \hline
Data & Peak position & Major axis \tablefootmark{a} & Minor axis \tablefootmark{a} & P.A. \tablefootmark{a} & Flux & Peak flux \\
 & R.A., Dec. (ICRS)& (mas) & (mas) & (deg) & ($\mathrm{\mu Jy}$) & ($\mathrm{\mu Jy\,beam}^{-1}$) \\
\hline
 1mm-high & {$04^{\mathrm{h}}19^{\mathrm{m}}58^{\mathrm{s}}.478, +27^\circ09'56''.85$} & $413 \pm 88$ & $351 \pm 84 $ & $133$ & $ 994 \pm 160$ & $208 \pm 28$ \\
 1mm-combine & {$04^{\mathrm{h}}19^{\mathrm{m}}58^{\mathrm{s}}.472, +27^\circ09'56''.73$} & Unresolved & Unresolved & - & $417 \pm 24$ & $448 \pm 15$ \\
 3mm & {$04^{\mathrm{h}}19^{\mathrm{m}}58^{\mathrm{s}}.490, +27^\circ09'56''.77$} & Unresolved & Unresolved & - & $47 \pm 25$ & $78\pm18$ \\
\hline
\end{tabular}
\tablefoot{
These results were obtained using the CASA task \texttt{imfit}. We note that we did not identify polarized structures for K04166 due to its complicated geometry.
\tablefoottext{a}{We list the convolved size with the beam in this table, because the polarized emission, especially that of K04169, is compact and weak, which can be spatially unresolved in the deconvolved size from the beam.}
}
\end{table*}

\end{appendix}

\end{document}